\documentclass[12pt,twocolumn]{iopart}

\usepackage{graphicx}  
\usepackage{multicol} 
\usepackage{lipsum}
\usepackage{color}
\begin{document}

\title[]{Electronic and magnetic properties of the graphene densely decorated with 3d metallic adatoms
}

\author{Ma\l gorzata Wawrzyniak-Adamczewska}

\address{Faculty of Physics, Adam Mickiewicz University, Umultowska 85, 61-614 Pozna\'n, Poland}
\ead{mwaw@amu.edu.pl}
\vspace{10pt}

\begin{abstract}
The electronic properties of graphene decorated with Ni, Co, Cu and Zn adatoms 
is studied with the density functional theory approach. Within the analysis the spin-orbit 
interaction is taken into account. We focus on the case when the indicated $3d$
metallic adatoms form a perfect, close-packed single-atomic layer above the graphene 
surface. The two configurations are examined, namely the
adatoms in the {\it on-top}, and the {\it hollow} positions on graphene. 
First, we verify that the metallic adatoms in the close-packed structure 
do not form a covalent bonds with the graphene substrate. However, due to the proximity 
of the metallic adatoms to the graphene, the charge transfer from the adatom layer 
to the graphene takes place, and in consequence the graphene becomes $n$-doped. 
The observed charge transfer results from the arising hybridization between the 
graphene $2p$ and transition metal $3d$ orbitals. The proximity of metallic adatoms
modifies the magnetic state of the graphene. This effect is especially pronounced for the 
decoration with magnetic atoms, when the magnetic moments on the graphene sublattices 
are induced. The analysis of the band structure demonstrates that the charge transfer, 
as well as the induced magnetism on graphene, modify the graphene electronic properties near 
high symmetry points, especially the Dirac cones. The presence of the metallic adatoms 
breaks graphene $K-K^{'}$ symmetry and splits the bands due to the exchange coupling. 
We show that for the {\it hollow} configuration the gap opening arises 
at the $K(K^{'})$-point due to the Rashba-like spin-orbit interaction, while in the case 
of the {\it on-top} configuration the energy gap opens mainly due to the staggered potential. 
We also mapped the parameters of an effective Hamiltonian on the results obtained with the density 
functional theory approach.
\end{abstract}

\section{Introduction}

Graphene is one of the most prominent currently studied two dimensional material \cite{geim}. 
This stable single layer of carbon atoms forms a honeycomb lattice and possesses 
extraordinary electronic properties, i.e. a linear band dispersion for low energies
near $K$ and $K^{'}$ symmetry points. The exciting electronic properties 
of graphene originate from the fact that its structure consists of the two 
equivalent sublattices, with the assigned pseudospins \cite{castro,novos1,kane}.

Since it is technologically inconvenient to obtain free standing graphene
and incorporate it into electronic devices, the graphene growth on various
substrates, as well as its decoration with various adatoms, or formation of
the hetero-layered structures addresses much attention \cite{Li}-\cite{ZhangP}. 
The structural incorporation of graphene with other materials may lead to the enhancement
of the spin-orbit interaction, which destroys the perfect massless-relativistic picture 
for charge carriers and introduces the gap in the energy spectrum 
as well as a spin-splitting of the bands. 
This feature is desirable, 
if addressing the graphene as a part of the electronic devices for spin-dependent 
transport, spin-filtering and magnetic valves. On the other hand, the unbounding 
of the graphene from substrates, e.g. by intercalation, in order to recapture its unique 
relativistic properties is also desirable. 

The magnitude of the intrinsic spin-orbit coupling in pristine graphene is small -- of orders 
of tens $\mu eV$ -- and originates from the graphene $d$-orbitals hybridization \cite{gmitra-konschuh}. 
Therefore, a scientific effort is done to study the r\^ole of substrate
or decoration with adatoms. First theoretical and experimental works analyzed 
the r\^ole of the metallic substrate on the electronic properties of graphene 
\cite{Gao}-\cite{Voloshina}, however the relativistic effects were not taken into account 
in those studies. Recently, several theoretical works appeared considering quantitatively 
the influence of the spin-orbit interaction on the symmetry breaking near $K$, $K^{'}$ points
and its influence on the modification of the linear dispersion of the Dirac cone for the graphene 
placed on the metallic surfaces \cite{Frank,Rader,Mertig,Zollner,Dyrdal2D}.
The properties of grahene decorated with metallic adatoms was also discussed \cite{Chan}. 

In this paper we focus on how the electronic properties of the graphene are modified when 
the graphene is decorated with a single layer of the close-packed $3d$ metallic adatoms.
The paper is organized as follows. First, the considered structures and applied methods are described. 
Next, taking into account the spin-orbit interaction as implemented in the applied pseudopotential DFT 
method, the general electronic properties, regarding charge transfer and the modification of the band 
structure of the graphene in the proximity of the metallic layer are considered. Subsequently, the insight 
into the orbital hybridization between the graphene and the metallic layer is presented. Then, 
the proximity induced magnetism on graphene is described. In the end, the quantitative
analysis of the influence of the spin-orbit interaction of the Rashba-like
type, the exchange coupling and the staggered potential at the Dirac point of the specified heterostuctures is
presented and discussed.  

\section{Method and System Geometry}

The presented results were obtained using a DFT approach as implemented in the plane-wave 
pseudopotential Quantum Espresso code \cite{QE}. We used fully-relativistic pseudopotentials 
in the Perdew-Burke-Ernzerhof (PBE) parametrization for the exchange--correlation functional \cite{PBE}. 
A plane-wave energy cutoff was set to 80 Ry, while charge density cutoff to 600 Ry, for all atomic species. 
The applied pseudopotntials contain the $2s$, $2p$ projections of valence states for carbon atoms, 
while $3d$, $4s$ and $4p$ projections for metallic adatoms. Within the analysis the Hubbard $U$ 
corrections for the $3d$ adatomic orbitals, in rotationally invariant scheme, were taken into 
account \cite{Liechten}. The value of the effective $U$ parameter for the nickel
was set to $U=6$ eV, while for the cobalt and zinc we assumed $U=2$ eV \cite{Co-U}, and for copper
we set $U=1$ eV. We also took into account the semi--empirical van der Waals corrections, since they are 
important for the large systems with dispersion forces \cite{grimme,barone}.

\begin{figure}[h]
\begin{center}
\begin{multicols}{3}
\includegraphics[scale=0.2,angle=0]{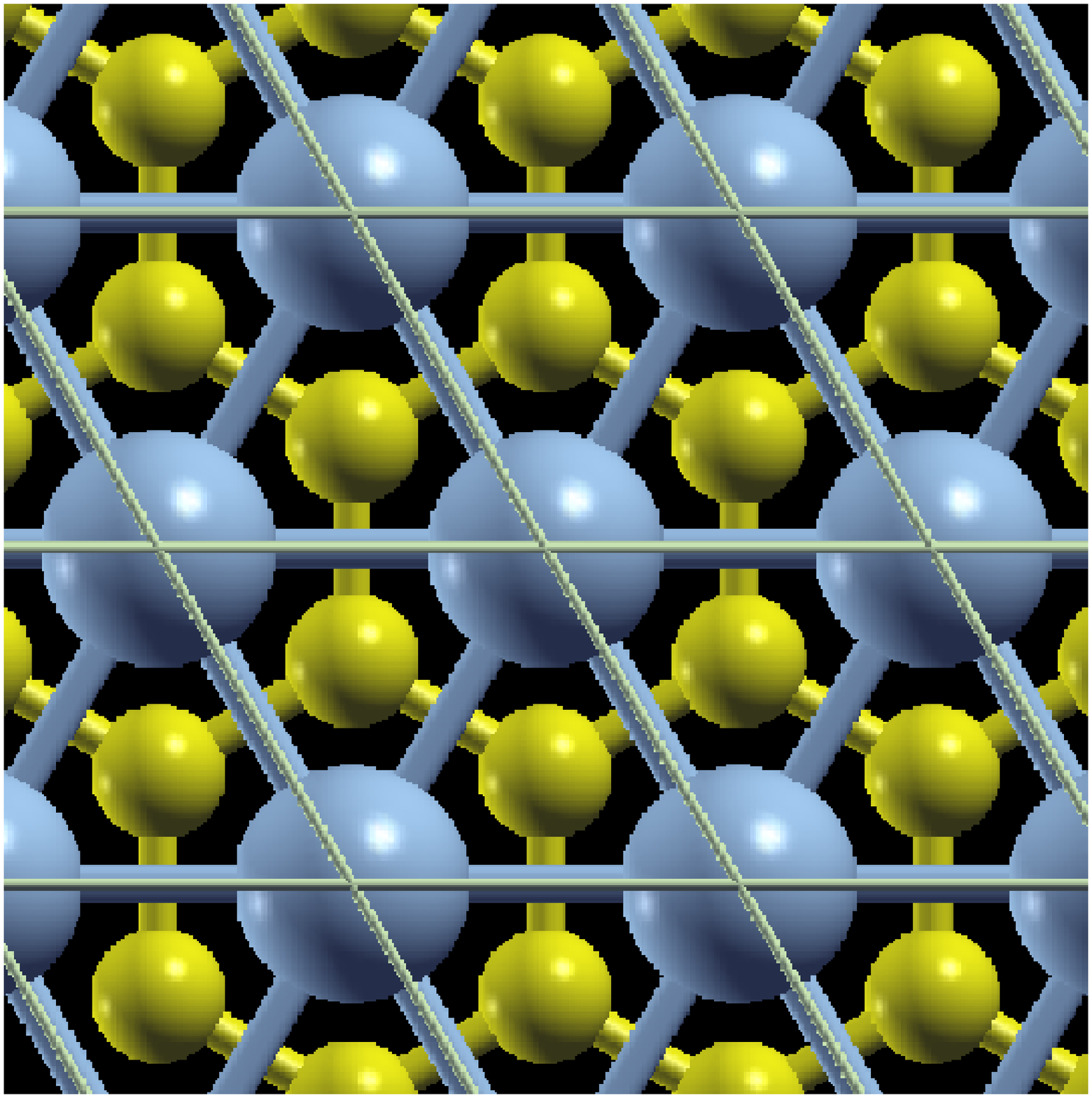}
\includegraphics[scale=0.2,angle=0]{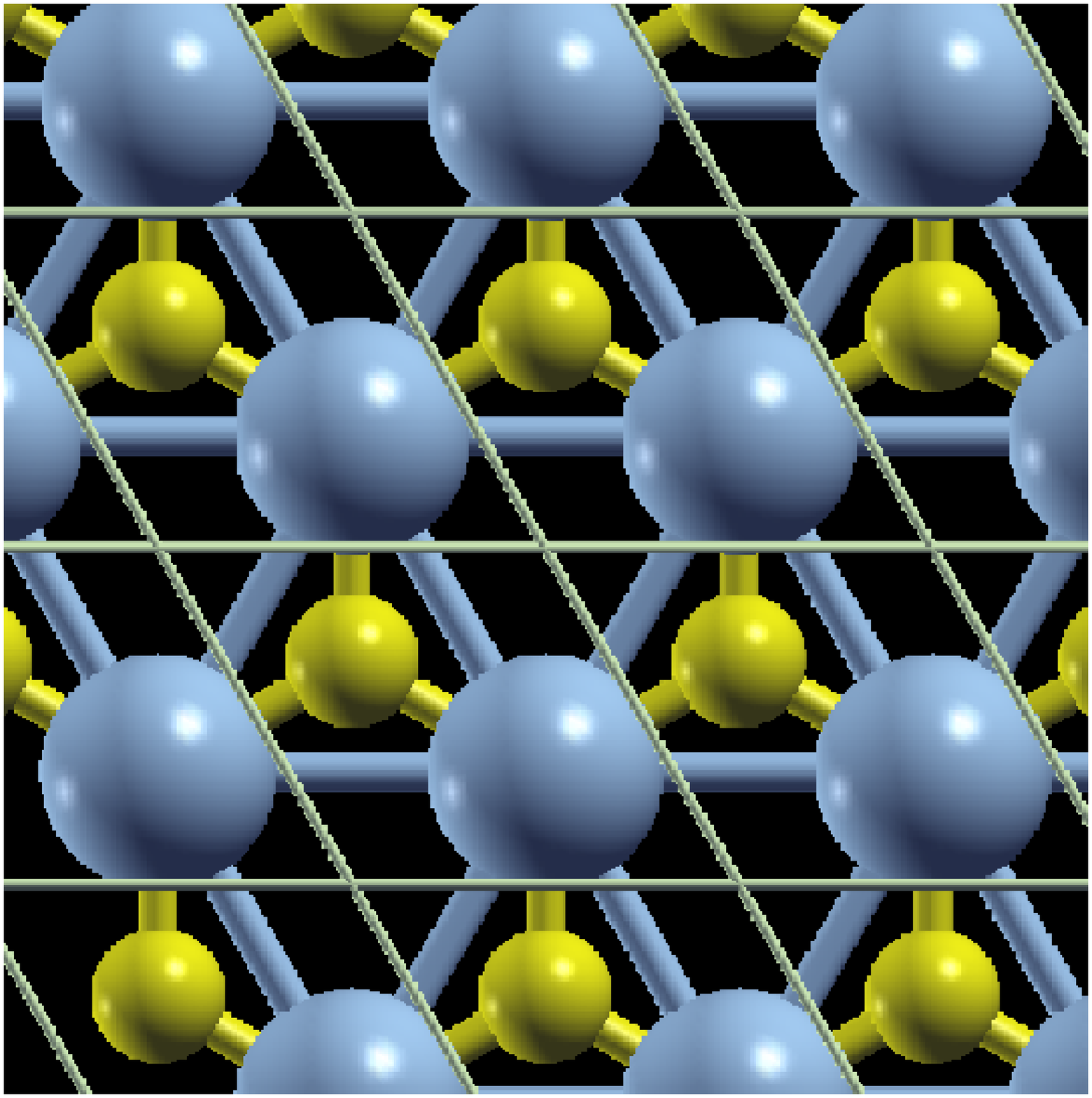}
\includegraphics[scale=0.35,angle=0]{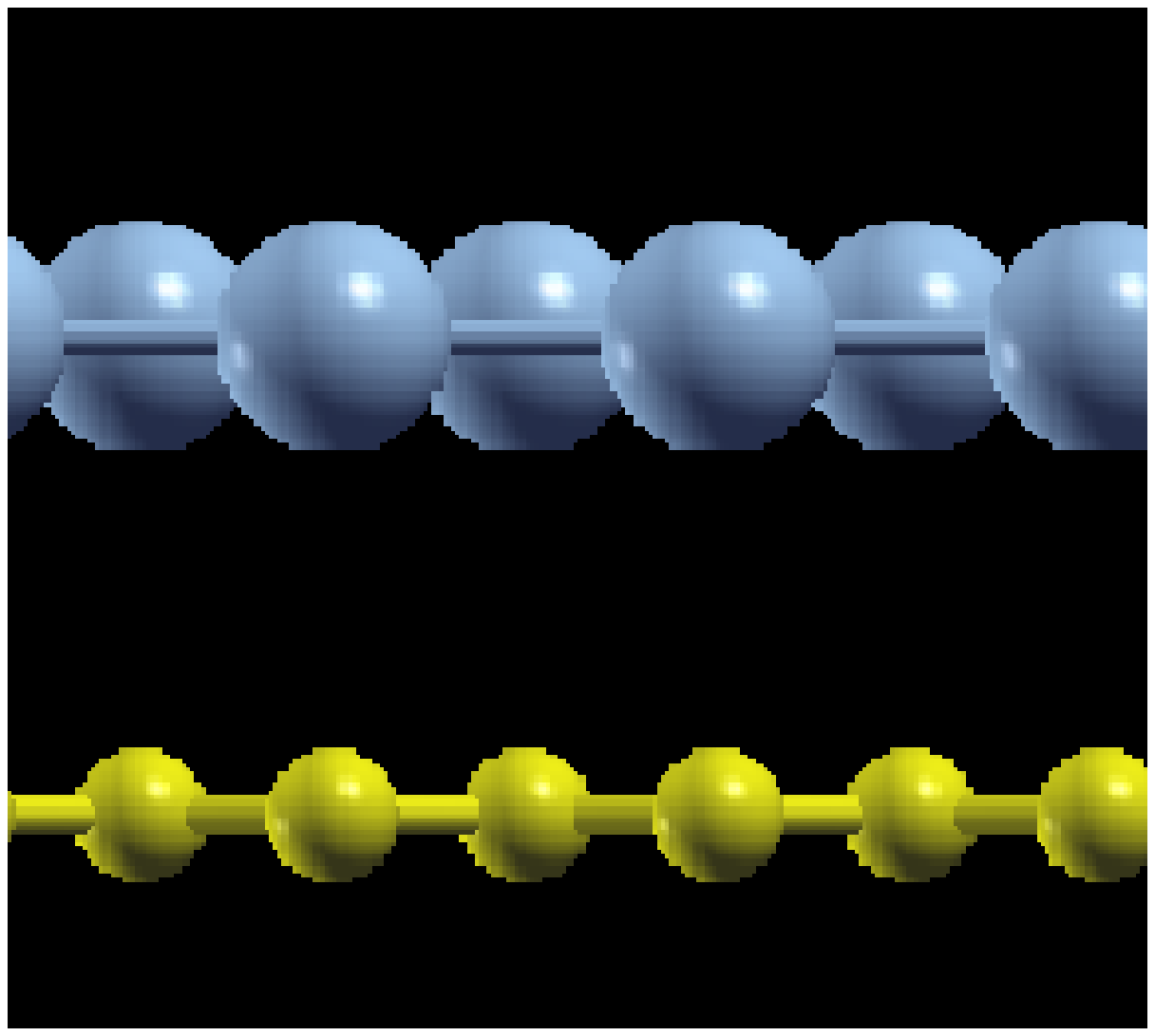}
\end{multicols}
\caption{\label{geo1}The top view of the graphene decorated with adatoms
in the {\it hollow} (left) and the {\it on-top} (middle) configurations. 
The specified adatoms (blue) form a close-packed layer above the graphene (yellow). 
The picture on the right hand side shows the side view 
of the two-layered structure in the {\it on-top} configuration. 
The graphene--metallic layer distance differs for adatom species and their 
positions on graphene, what is shown in Tab.\ref{tab1}.}
\end{center}
\end{figure}

The considered hexagonal supercells consists of the two graphene atoms, of A and B sublattices, 
and the metallic adatom in the {\it on-top} or {\it hollow} positions. The geometries of the considered 
structures with indicated supercellls are presented in Fig.\ref{geo1}. The periodic slabs were 
separated by 12 \AA~of vacuum. We used a uniform 30$\times$30$\times$1 Monkhorst-Pack k-mesh \cite{MKP} 
to sample the first Brillouin zone for the hexagonal supercells. Within the applied parameters and 
pseudopotentials, the optimized C--C distance of pristine graphene is 1.42
\AA.

Since we focus on the close-packed decoration, the choice of the specified adatoms is 
motivated by the adequate matching of the lattice constants of graphene and the lattice 
constants of the established metallic layers. We focus on the studies of decoration with 
nickel and copper atoms, since the Ni--Ni interactomic distance on the Ni(111) surface 
is 2.48 \AA~and the Cu--Cu interatomic distance on Cu(111) surface 
is 2.55 \AA, as well as with cobalt and zinc atoms, since the interatomic 
distances for the Co(0001) and Zn(0001) close-packed surfaces are 2.51 \AA~and Zn--Zn 2.66 \AA, 
respectively \cite{Ni-lattice,CoCu-lattice}. 
The optimized distances between the graphene and the close-packed adatomic 
layers are given in Tab.\ref{tab1}. The presented values suggest the physisorbtion 
of the $3d$ metallic adatoms on the graphene substrate. This statement is then confirmed 
by the analysis of the binding energy calculated per grahene C--atom. The values of 
the binding energy, presented in Tab.\ref{Ebin}, are far lower than 500 meV, what allows 
us to conclude that no covalent bounds are formed between the graphene and the close-packed 
adatoms. 
\begin{table}
\caption{The values of the optimized graphene -- adatom layer distances d
and in \AA, for the {\it on-top} (t) and {\it hollow} (h) configurations.
\label{tab1}}
\begin{center}
  \begin{tabular}{  l  c  c  c  c  c  c  c  c  c  c }
    \hline \hline
               & Ni(h)  & Ni(t)  & Co(h)  & Co(t)  &  Cu(h)   & Cu(t)  & Zn(h)  & Zn(t)  \\ \hline 
            d  & 3.3754 & 3.0648 & 3.2112 & 3.1218 &  3.0787  & 3.1011 & 3.4636 & 3.3154  \\ 
    \hline \hline
  \end{tabular}
\end{center}
\end{table}

The values of the total energy for the graphene with metallic adatoms is shown in Tab.\ref{Etot}. 
From the analysis of the total energies in the specific configurations it is clearly seen, that 
the {\it on-top} configuration is the preferred one for nickel and cobalt layers, namely for
the ferromagnetic metallic layers. 
The difference in total energy between the {\it hollow} and {\it on-top} configurations is
of order of 14 and 20 meV for the nickel and cobalt layers, respectively. This suggests that, 
while close-packed, the mentioned atoms do not tend to nest the {\it hollow} position 
on the graphne lattice and rather be imbedded in the {\it on-top} position. The difference of 
the total energies for the copper layers is of order of few $\mu$eV merely, in favor of 
the {\it on-top} configuration, while the zinc atoms prefer to nest the {\it hollow} position, 
and the difference between the total energy in the {\it on-top} and {\it hollow} configurations
for the latter adatom species is found to be equal 2.5 meV.

At this point we emphasize however, that supercell we applied to our DFT
analysis is too small to include the effects of the Moire patterns of the
graphene--metal interface \cite{Murata,Colomer,Moire-Ni,Moire-Pt}, as well as the buckled graphene patterns
\cite{Deng}. Hence, the analysis of the spatially extended system, with larger
supercells, could led us to different conclusions regarding the structure
privileges. Moreover, we verified that the reduction by half of the adatoms concentration 
on the graphene significantly changes the bonding type. Namely, when the adatoms are spread 
on graphene at lower concentrations they tend to form covalent bonds with 
the graphene substarate.

\begin{table}
\caption{The values of the total energies in Ry for the {\it on-top} and {\it
hollow} configurations of the graphene decorated densely with the $3d$
metallic adatoms.
\label{Etot}}
\begin{center}
  \begin{tabular}{  l  c  c  c  c }
    \hline \hline
       & Ni              &           Co   &    Cu           & Zn               \\ \hline
on-top & -124.29904316   & -99.88880617   &  -141.30003430  & -172.23111898    \\
hollow & -124.29798240   & -99.88733592   &  -141.30002475  & -172.23130580    \\
    \hline \hline
  \end{tabular}
\end{center}
\end{table}

\begin{table}
\caption{The values of the binding energies in eV for the {\it on-top} and {\it
hollow} configurations of the graphene decorated densely with the $3d$
metallic adatoms. 
\label{Ebin}}
\begin{center}
  \begin{tabular}{  l  c  c  c  c }
    \hline \hline
             & Ni     &     Co & Cu    & Zn     \\ \hline   
   on-top    & 133.34 & 107.66 & 62.82 & 103.94 \\ 
   hollow    & 117.45 &  87.74 & 62.69 & 106.49 \\
    \hline \hline
  \end{tabular}
\end{center}
\end{table}

\section{Results}

\subsection{Band structure}

In order to verify the results with spin-orbit coupling, first we calculated the intrinsic spin-orbit 
induced gap for pristine graphene at the K-point. With the established DFT parameters, supercells 
and pseudopotentials we obtained value of the intrinsic spin-orbit gap at 10 $\mu eV$. This value stays 
in excellent agreement with the previous DFT results \cite{gmitraTB} and also gives a precision of our 
spin-orbit induced energy gap calculations that are presented in the final part of this work.

\begin{figure}[h]
\begin{center}
\begin{multicols}{2}
\includegraphics[scale=0.60]{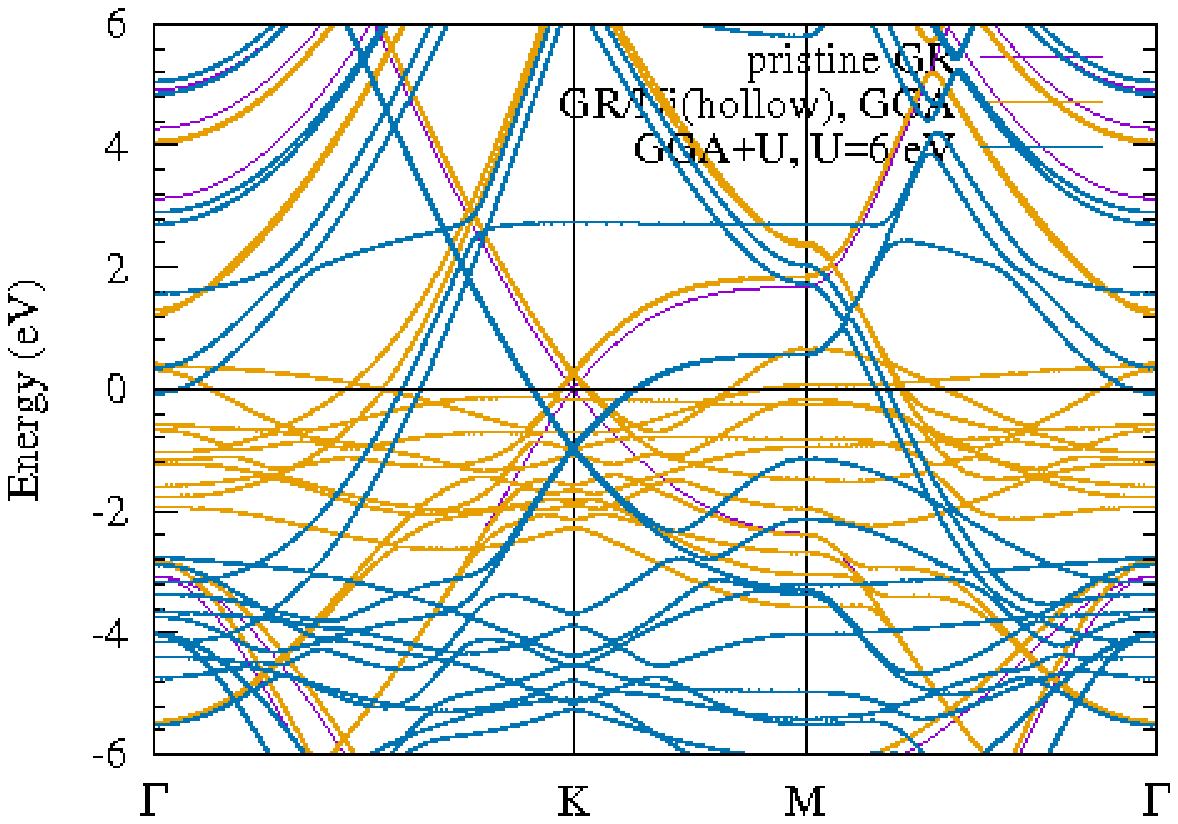}\\
\includegraphics[scale=0.60]{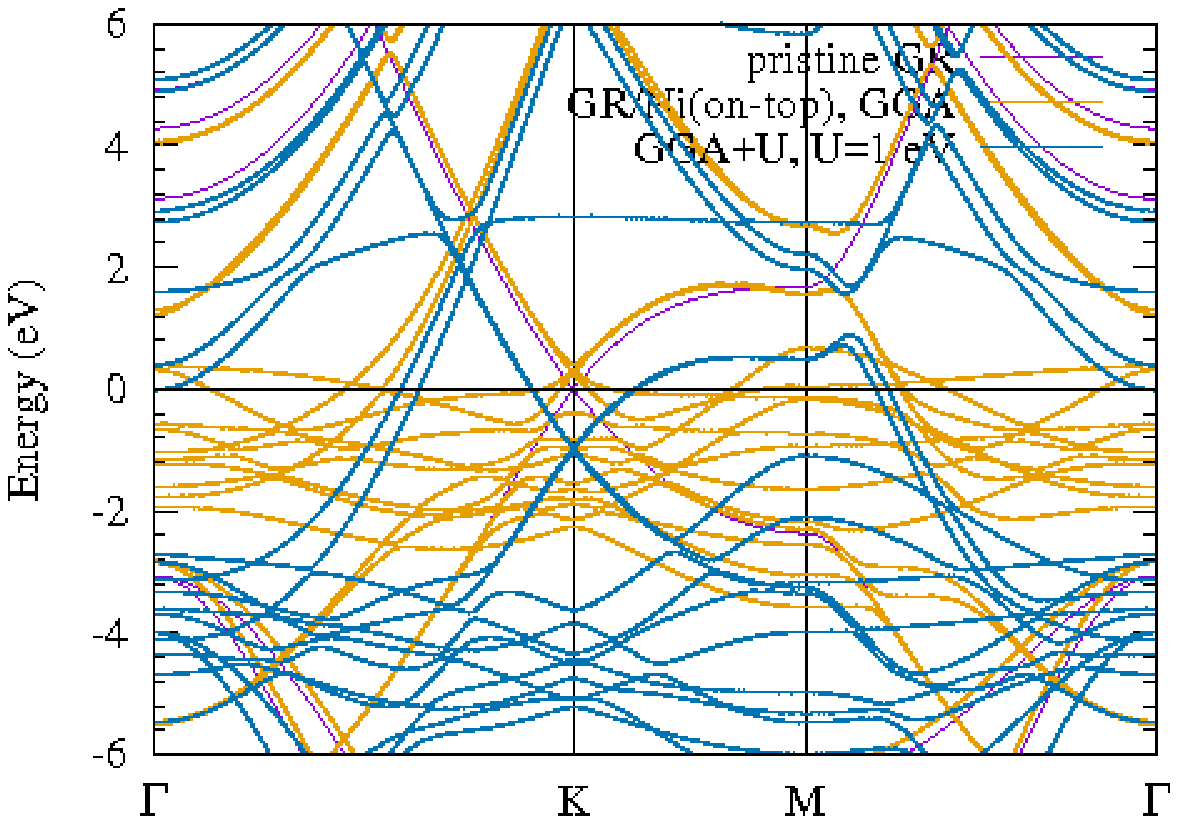}\\
\includegraphics[scale=0.60]{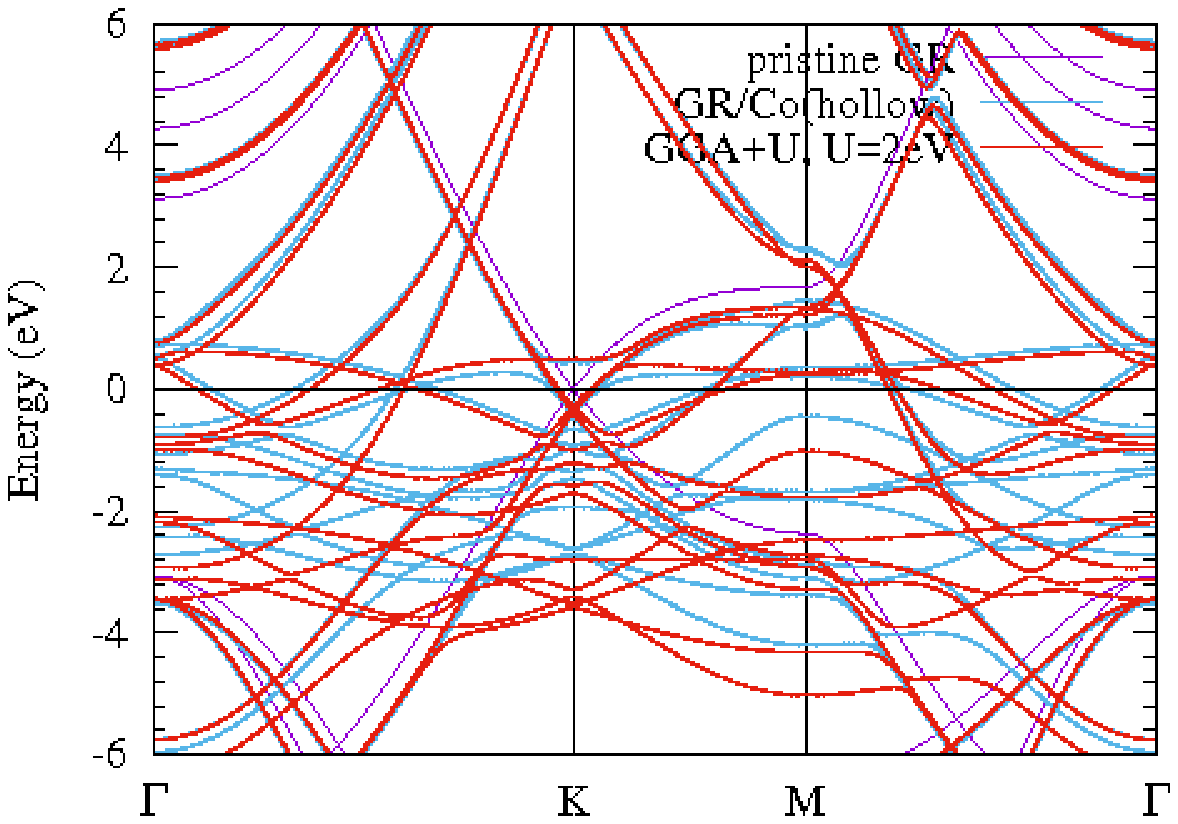}\\
\includegraphics[scale=0.60]{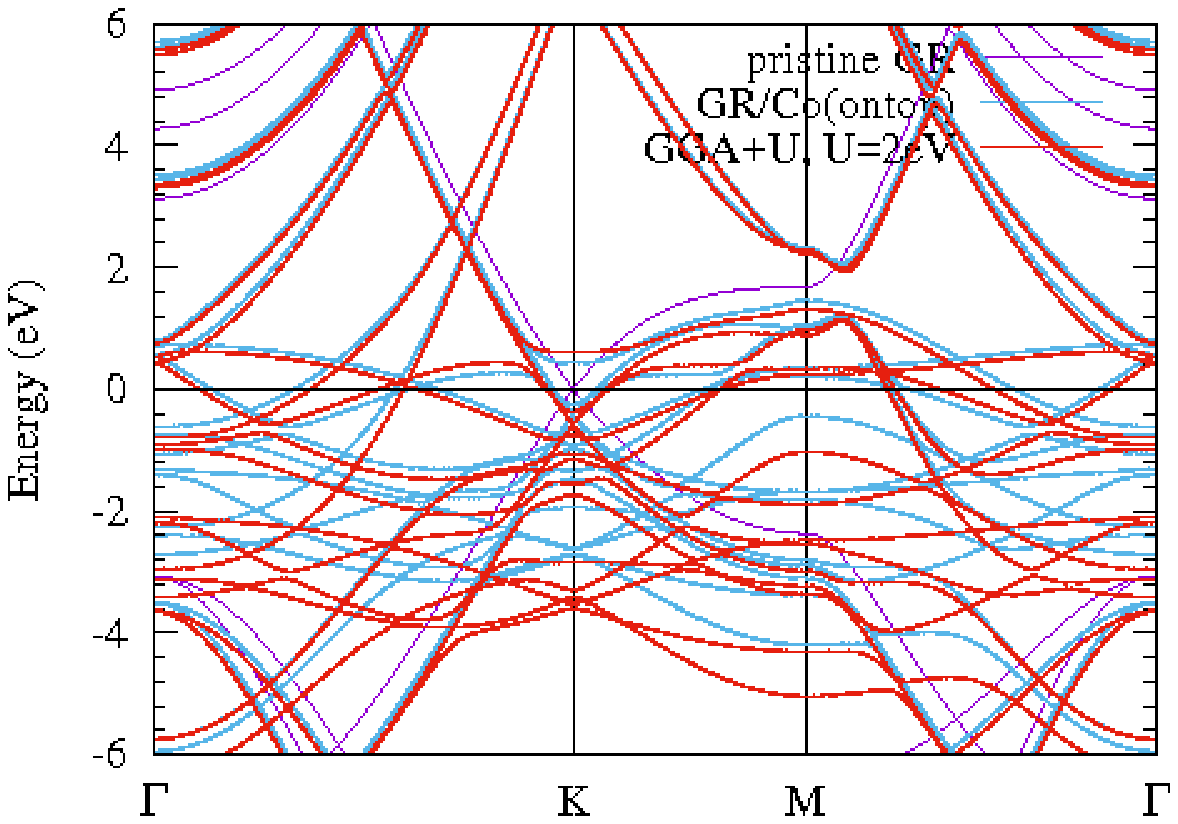}\\
\includegraphics[scale=0.60]{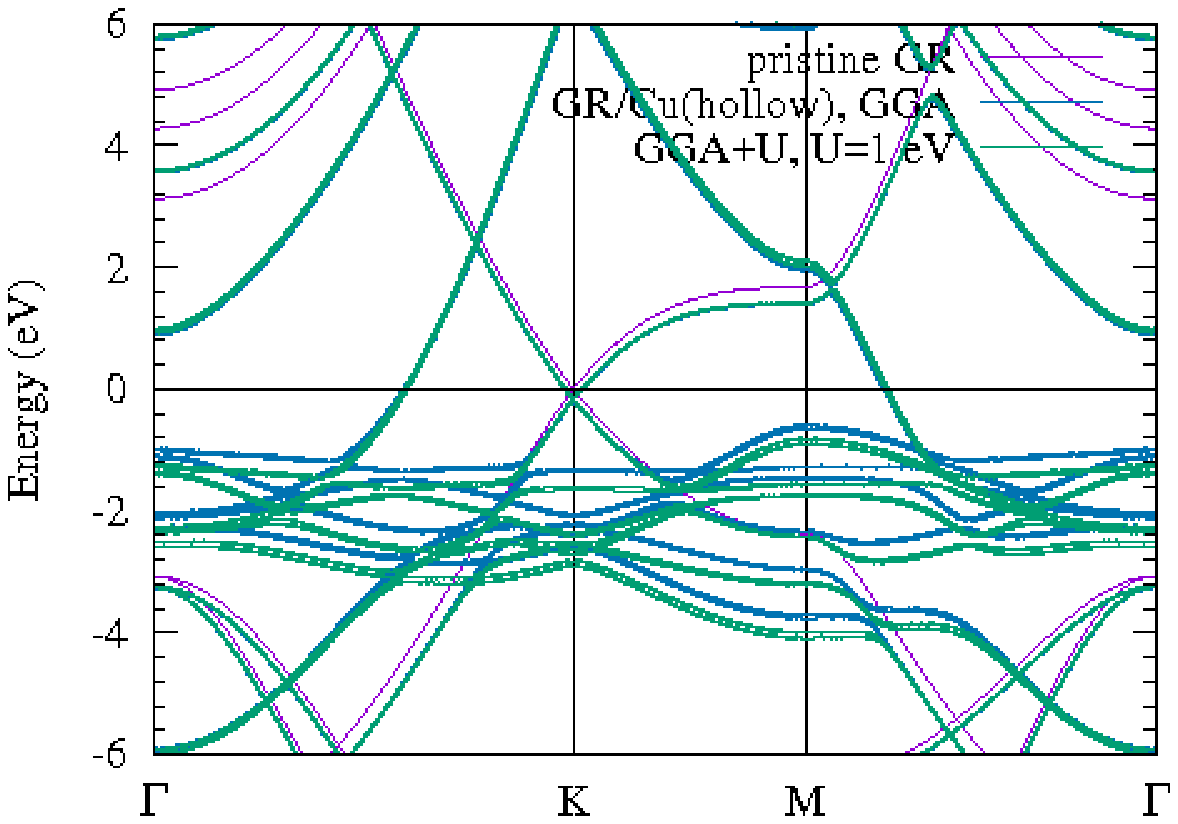}\\
\includegraphics[scale=0.60]{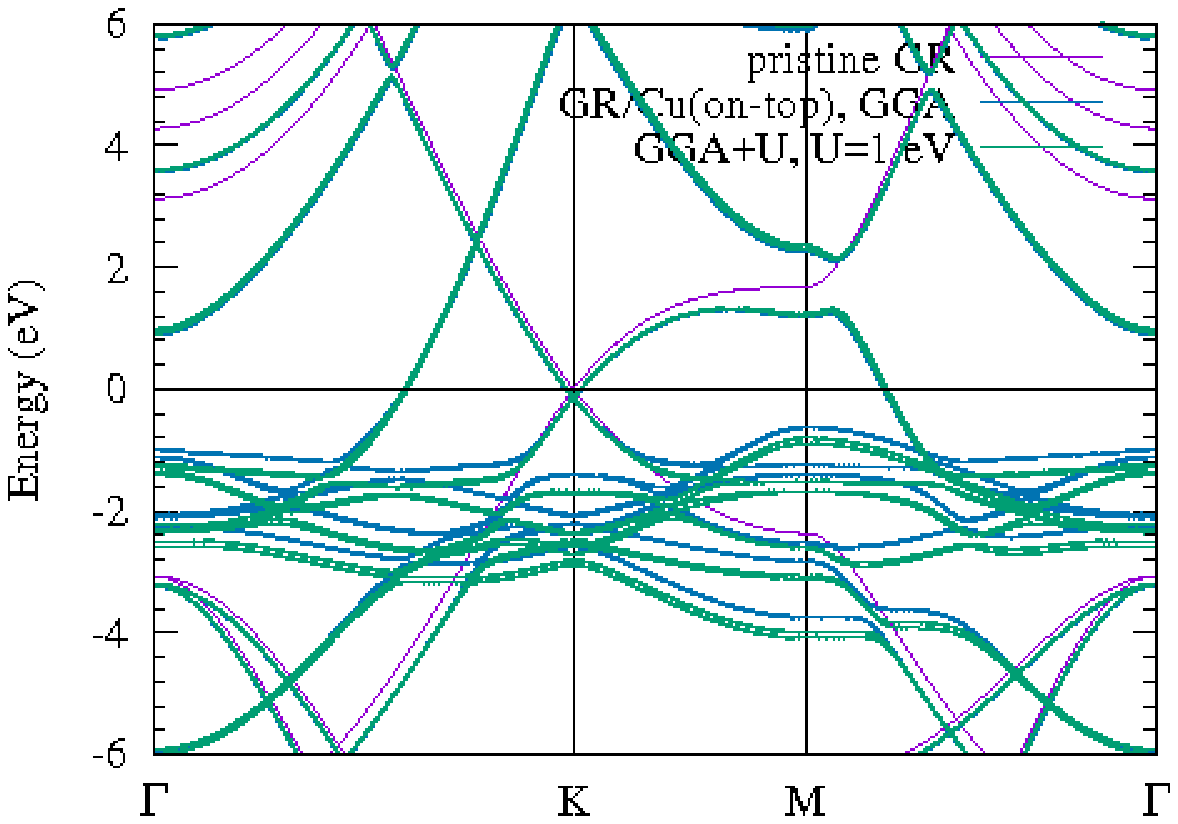}\\
\includegraphics[scale=0.60]{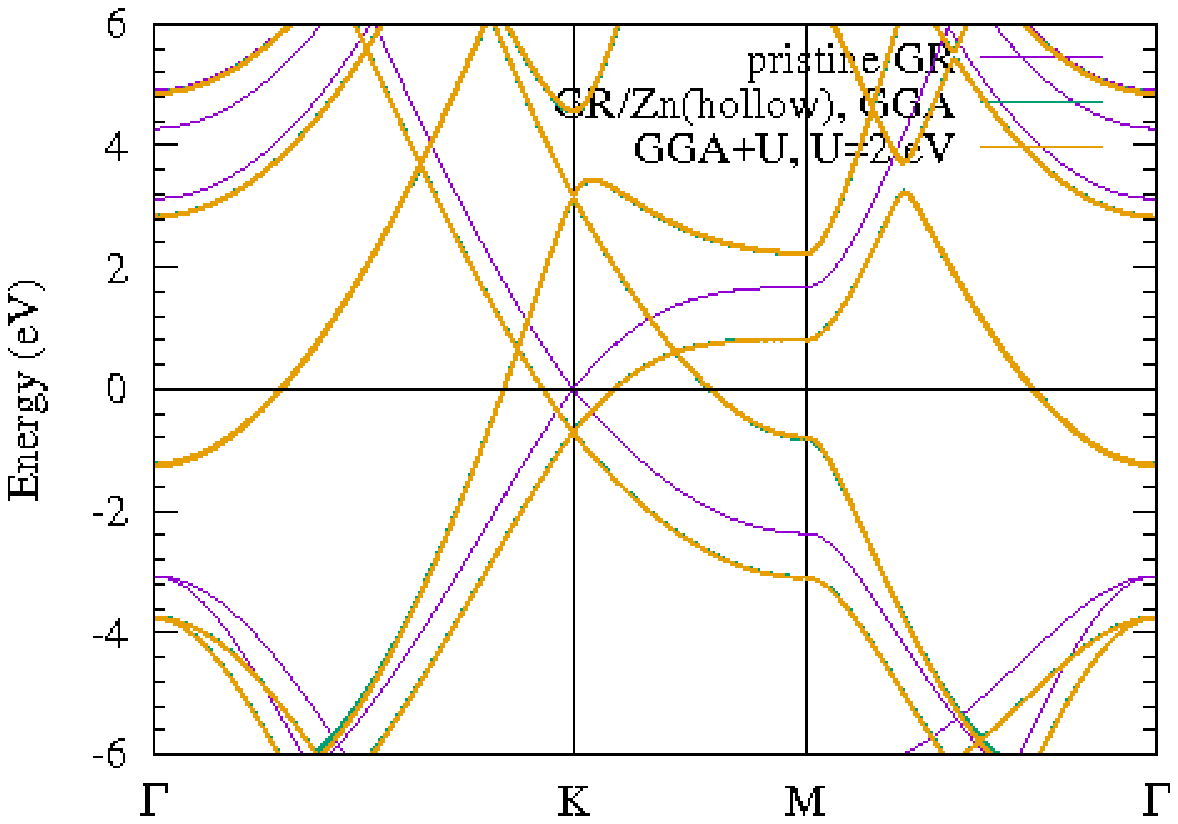}\\
\includegraphics[scale=0.60]{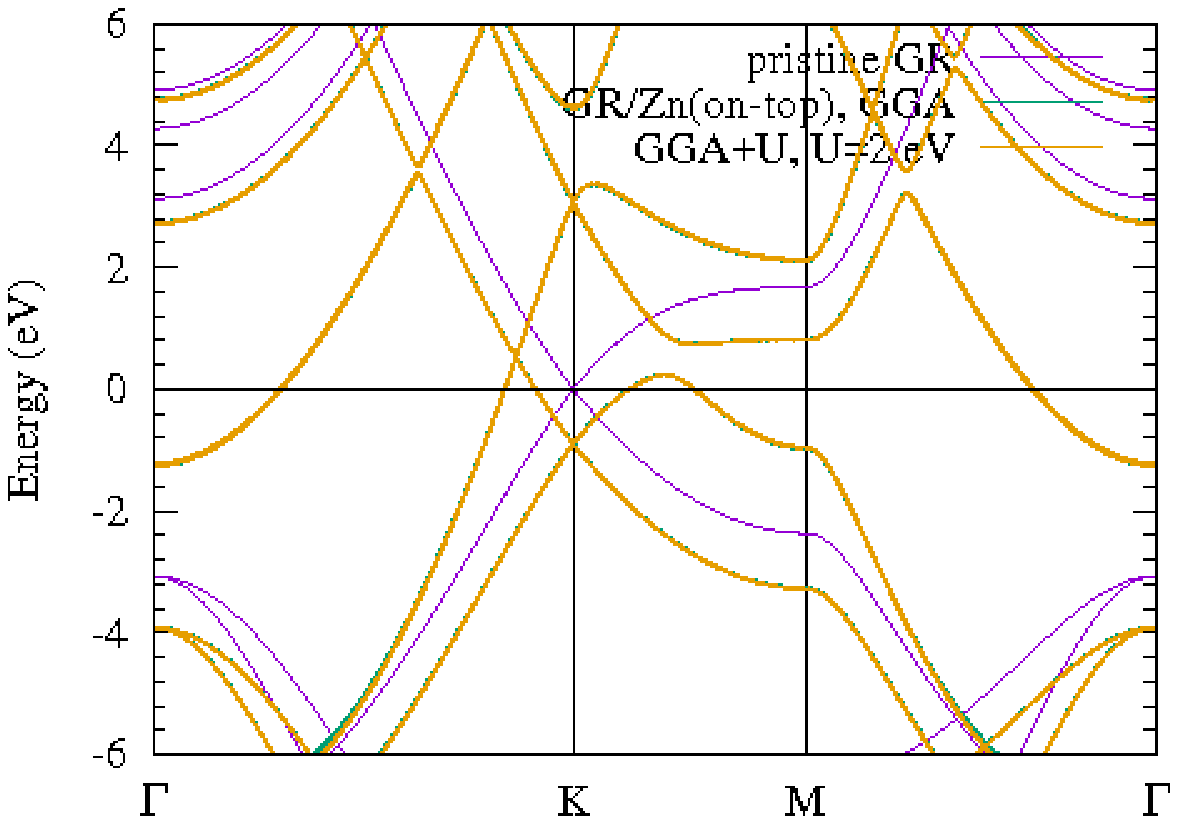}\\
\end{multicols}
\caption{\label{hbt} The band structures for the graphene with close-packed metallic
layers. The {\it hollow} and {\it on-top} configurations are specified in the figures. The band 
structures were calculated for the GGA only, or with additional assumption of the finite value of 
the Hubbard $U$ parameter, as specified in the figure. The band structure for the pristine graphene 
is plotted with the purple line for compartion.}
\end{center} 
\end{figure}

In Fig.\ref{hbt}, the band structures for the graphene densely decorated with $3d$ metallic adatoms 
for the {\it hollow} and the {\it on-top} positions are shown. In the {\it on-top} case the metallic 
adatom is situated above the graphene carbon atom of the A sublattice, hence a coupling with the A 
sublattice is assumed to be stronger than with the B sublattice. For the {\it hollow} configuration 
the equal couplings with both sublattices are present. The adatom type and position on the graphene 
lattice is indicated in the figure. The band structure 
for the pristine graphene is plotted with the purple line and presented in all figures for compartion. 
The figures show the course of the band structures obtained in the calculations where the Hubbard $U$ 
corrections of the $3d$ metallic orbitals are or are not taken into account, as indicated. For the presented 
energy range, there is no much difference in the band structure between the {\it hollow} and the {\it on-top}
configurations along reciprocal path for the particular considered $3d$ adatomic layer. The exception 
is the vicinity of the $M$-point, where the difference in the band structure for the conduction states 
between the {\it hollow} and the {\it on-top} cases is pronounced. 
This difference is especially pronounced for Ni, Co and Cu adatomic layers around 2 eV, and appears 
despite the treat of the Hubbard $U$ corrections.
The difference in the band structure course near the $M$-point 
is attributed to the fact that for the {\it on-top} configuration the symmetry 
breaking appears in the reciprocal space near $M$-point on $M$--$\Gamma$ path.

At the presented energy range, the linear dispersion near $K$ symmetry point 
is reproduced for all considered adatoms. However, the shift in energy of the Dirac 
point with respect to the Fermi Energy $E_{\rm F}$, here denoted by $E_{\rm D}$, appears. 
We attribute this shift to the strength and type of the hybridization between the graphene 
and adatomic layer. The $n$-doping type is observed for all studied adatoms, since the 
Dirac point of the considered systems is shifted below the Fermi energy. It is worth 
to underline, that taking into account the Hubbard corrections in the case
of nickel layer is crucial, since it alters the graphne doping type, while for 
the zinc adatomic layer the modification in the course of the band structure introduced 
by the Hubbard correction is unnoticeable in the presented energy range, despite
the assumption of a quite large value of the $U$ parameter. 
In general, the two groups of the adatoms among the studied ones may be distinguished. There 
are the adatoms that tend to hybridize strongly with the graphene, and for them the shift of 
the Dirac point is significant -- these are nickel, cobalt and zinc, and the adatoms that can 
be regarded as weakly coupled to the graphene -- the copper. The values of the shift in energy of
the Dirac point $E_{\rm D}$ for all considered adatoms are presented in Tab.\ref{ED}. 
Our observation of the graphene doping type stays in agreement with the previous studies 
of the graphene placed on the metallic substrates \cite{Voloshina,kelly}.    

\begin{table}
\caption{The values of shift of the Dirac point, $E_{\rm D}$ in eV, for the indicated
$3d$ adatoms in the {\it hollow} and {\it on-top} configurations. 
\label{ED}}
\begin{center}
  \begin{tabular}{  l  c  c  c  c }
    \hline \hline
        &  Ni       & Co        &   Cu      & Zn        \\ \hline
 on-top & -0.9710 & -0.5869 & -0.1425 & -0.9059 \\
 hollow & -0.9824 & -0.3329 & -0.1718 & -0.7170 \\
    \hline \hline
  \end{tabular}
\end{center}
\end{table}

One may also observe, that the set of flat doping bands builds up below or/and around the Fermi energy. 
The position and spread in energy of the set of the doping bands depends on the adatom type. 
For the presented energy window the set of doping bands is clearly visible for the nickel, 
cobalt and copper adatomic layers, while for the zinc the doping bands lay below the presented 
energy frame. Importantly, for the nickel and cobalt adatoms, and lesser for the copper layer, 
the location in energy of the doping bands depends on the strength of the Hubbard $U$ corrections. 
In general, taking into account the Hubbard $U$ corrections results in dragging down
in energy of the valence bands, while the conduction states are less affected by these corrections. 
In the case of nickel, cobalt and to some extend copper, the increase of the $U$ correlations 
enhances the shifting down the doping bands. As mentioned above, the inclusion of the Hubbard 
correction not only influences the position of the doping bands, but also the position in energy 
of the Dirac point. Here, the two types of the adatoms may be extracted. The adatoms for which 
the Hubbard correction notably shifts down the Dirac point and/or the position of the doping bands --
for this group the adatoms forming the ferromagnetic layers, namely nickel and cobalt atoms 
are included. The other group consists of the adatoms, for which the Hubbard $U$ 
corrections do not influence the position of the Dirac point and the doping bands are merely moved. 
The later group represents the copper and zinc, namely nonmagnetic adatoms.  

The effect responsible for the shifting down the band structure, and in particular, the Dirac point 
below the Fermi energy is the charge transfer that takes place from the metallic layer to the graphene. 
This charge transfer is possible since the hybridization between the specific graphene and adatomic layer 
orbitals develops. 
The $n$-type doping for graphene on metallic surfaces is predicted theoretically 
\cite{Frank,yamamoto} and reported in experiment \cite{Dedkov}. 

\subsection{The $p-d$ hybridization}

\begin{figure}
\begin{multicols}{3}

\includegraphics[scale=0.45]{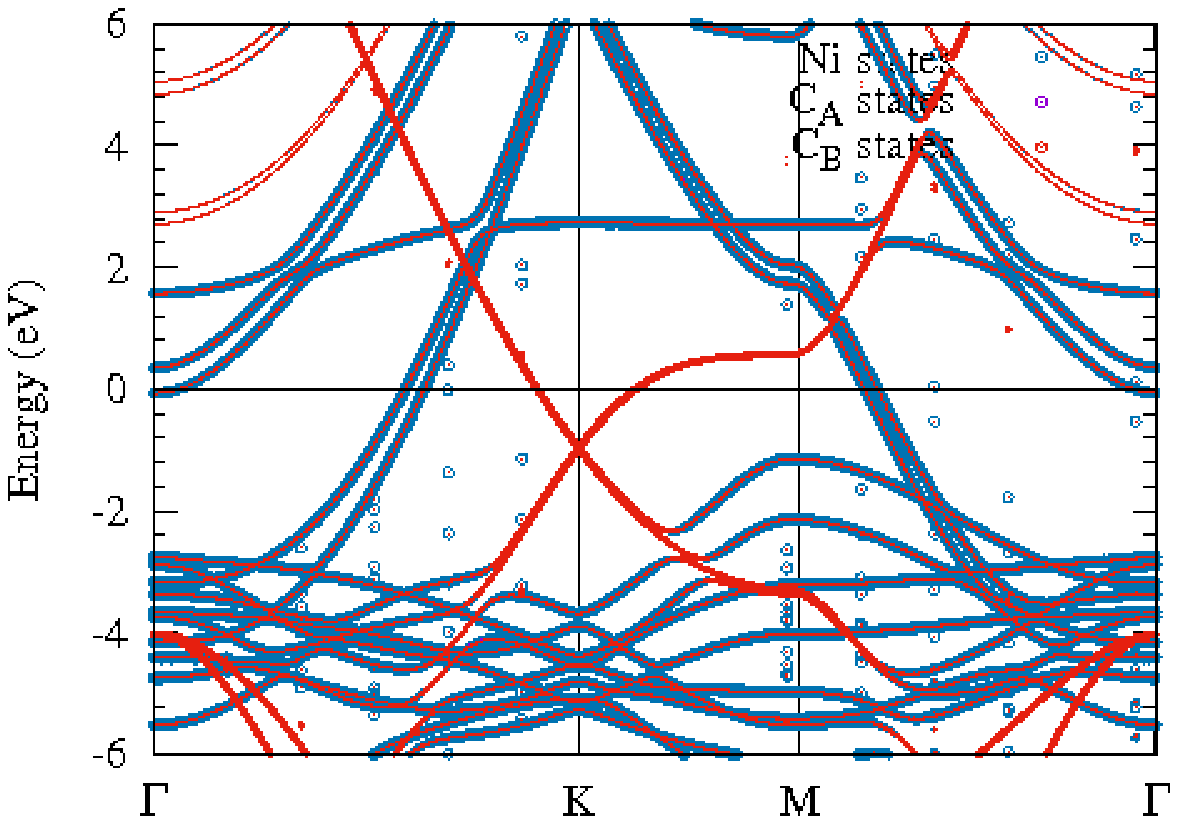}\\
\includegraphics[scale=0.45]{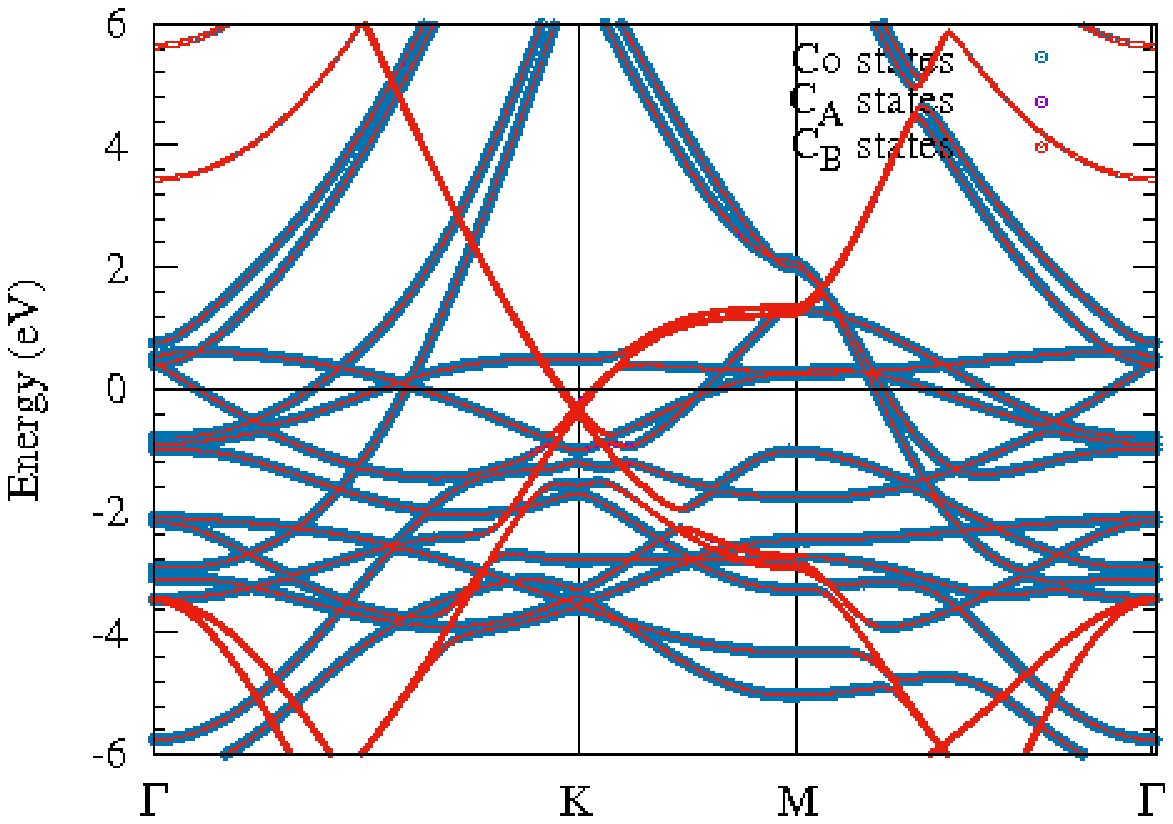}\\
\includegraphics[scale=0.45]{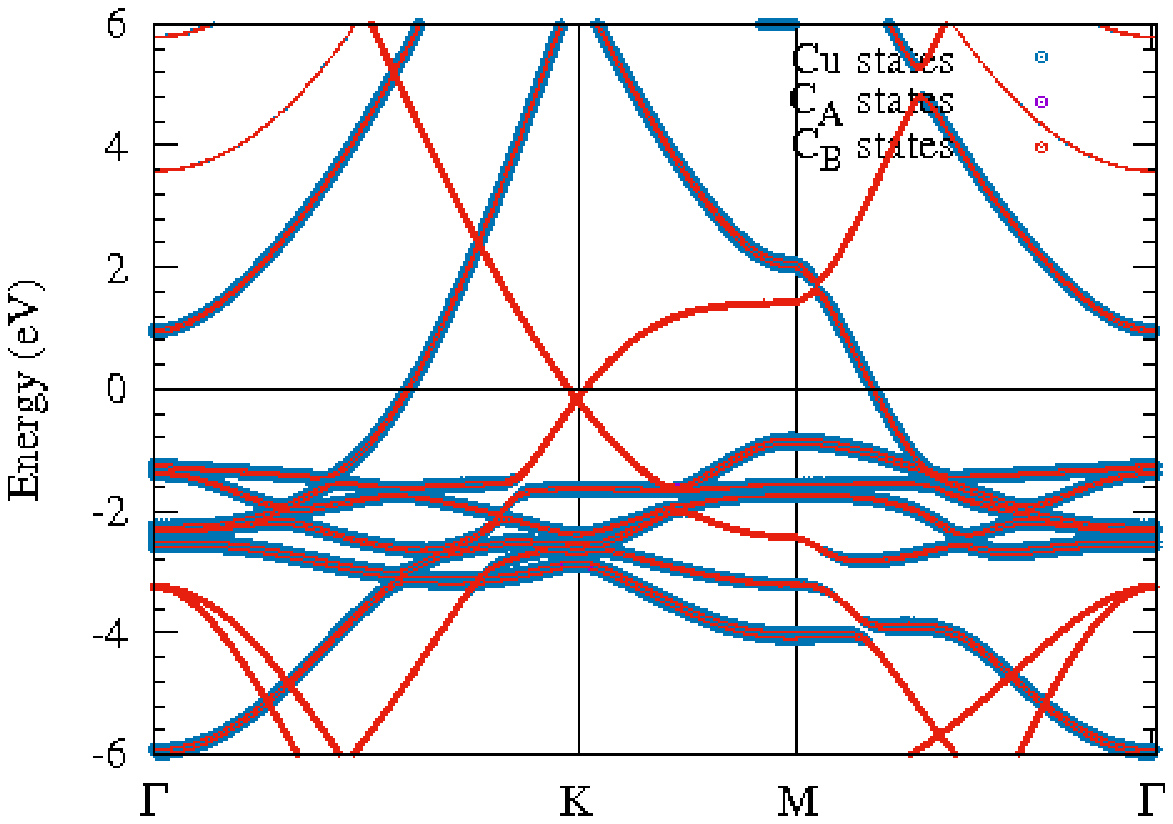}\\
\includegraphics[scale=0.45]{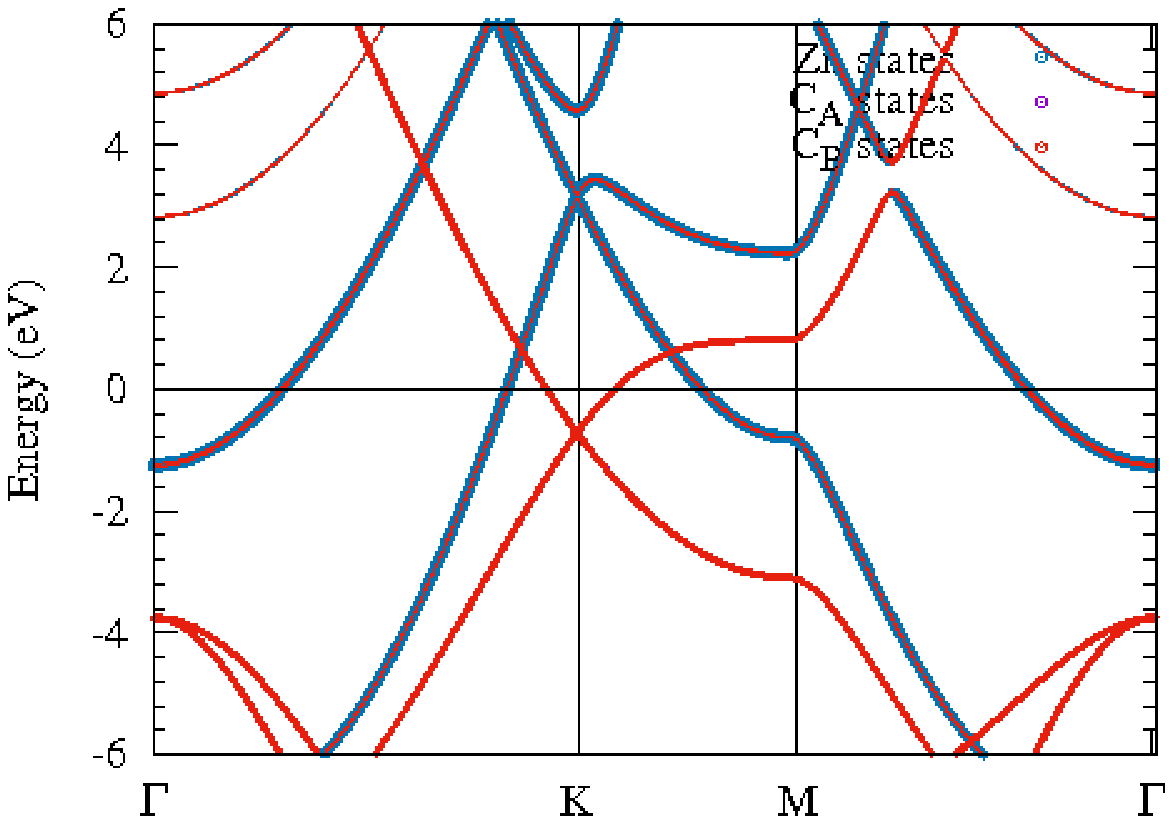}\\
\includegraphics[scale=0.45]{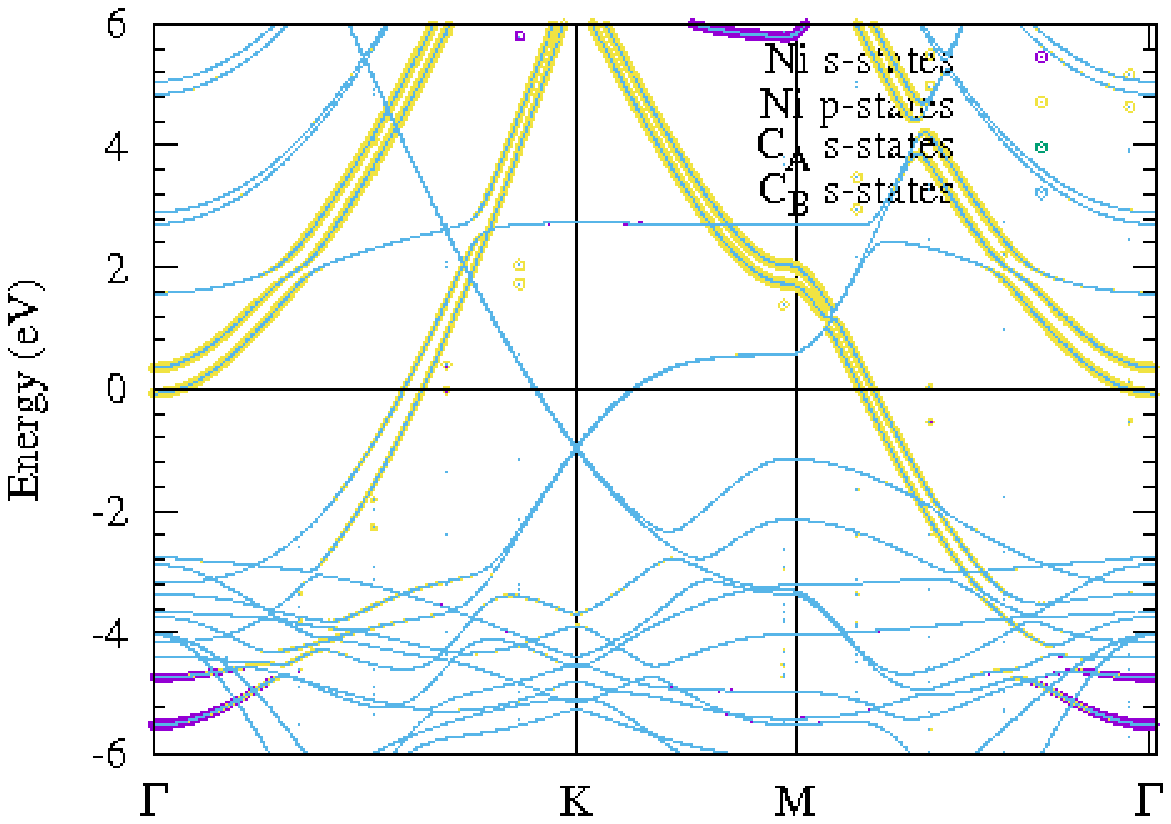}\\
\includegraphics[scale=0.45]{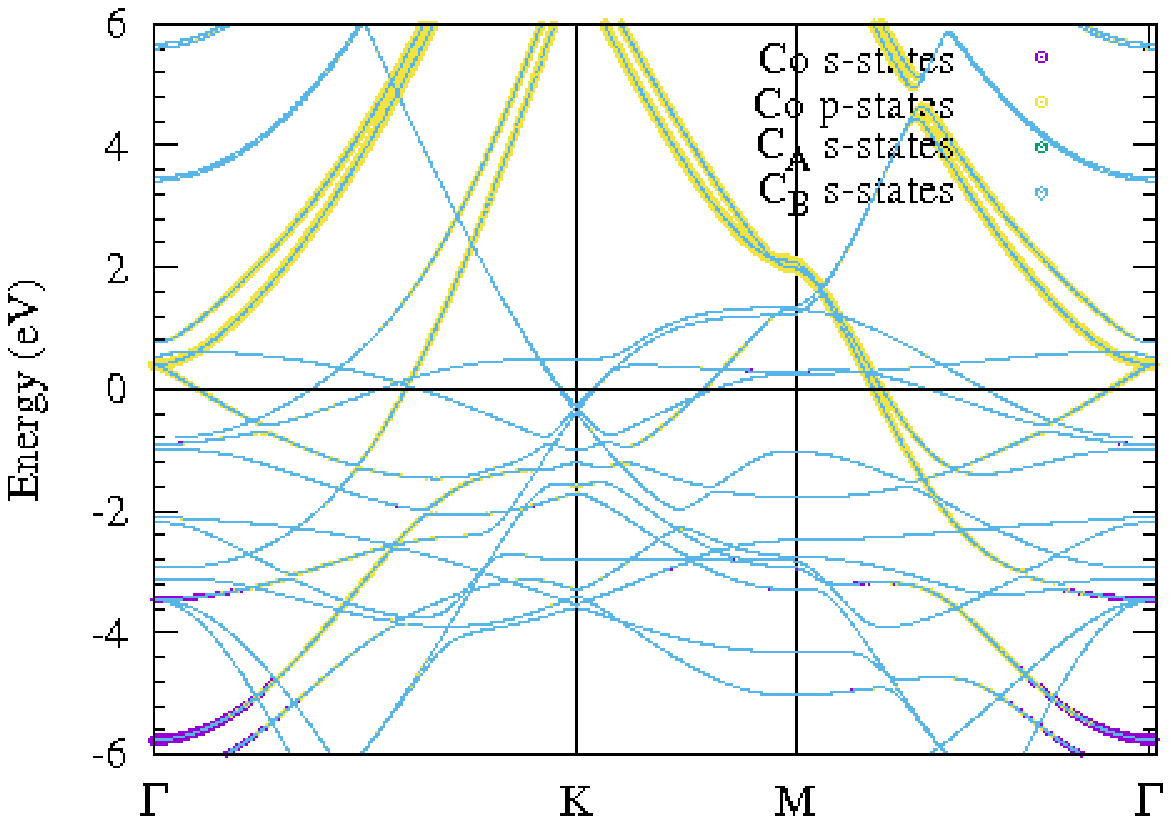}\\
\includegraphics[scale=0.45]{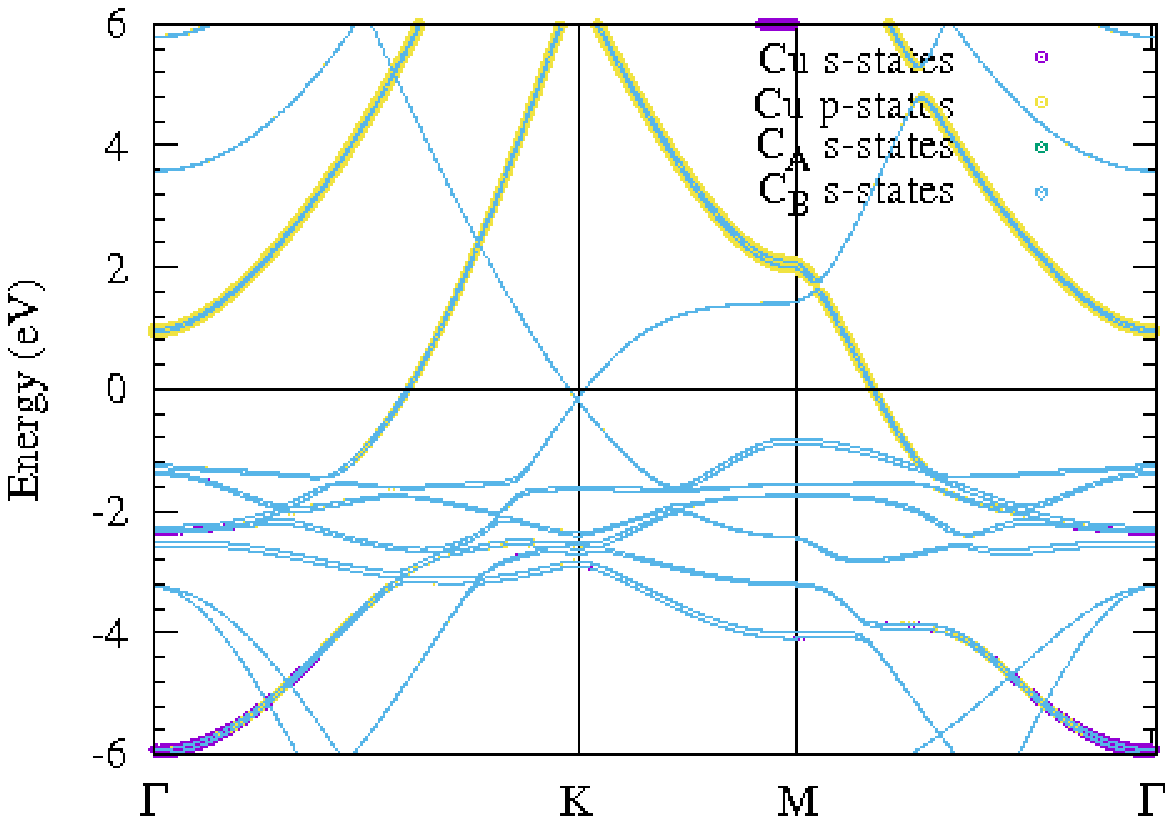}\\
\includegraphics[scale=0.45]{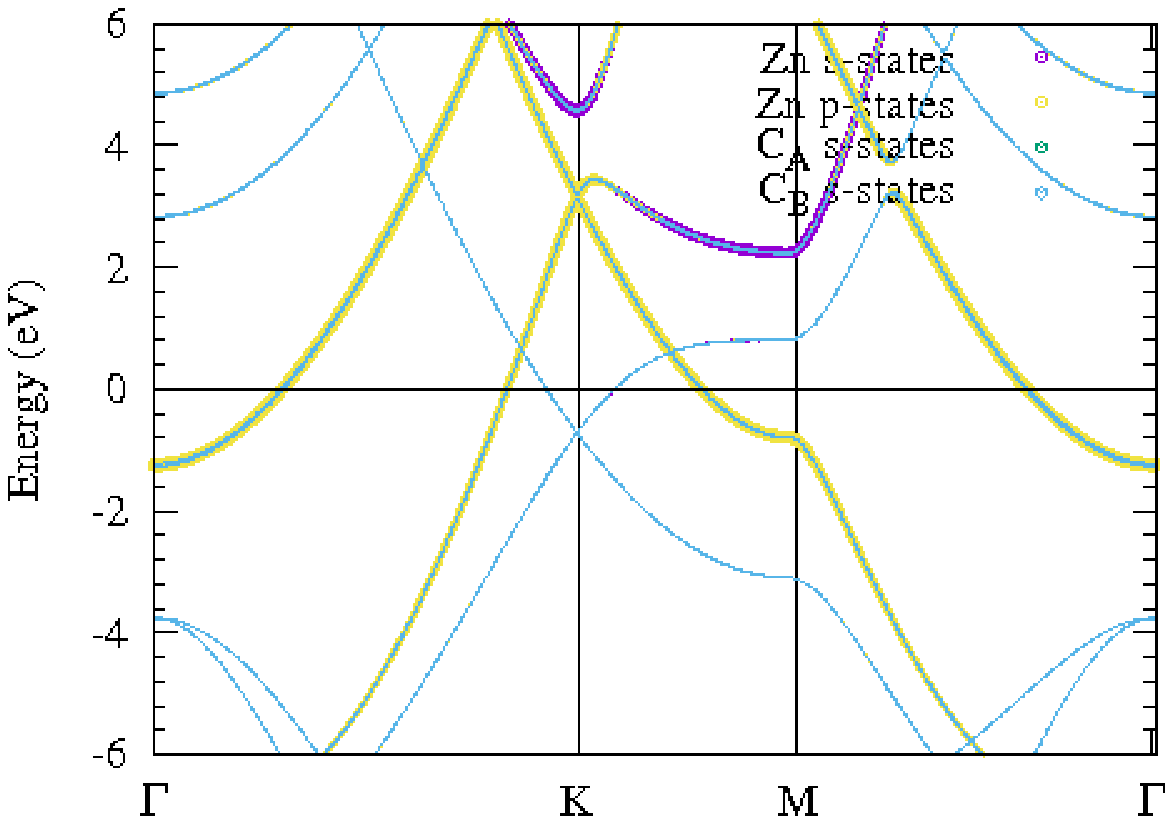}\\
\includegraphics[scale=0.45]{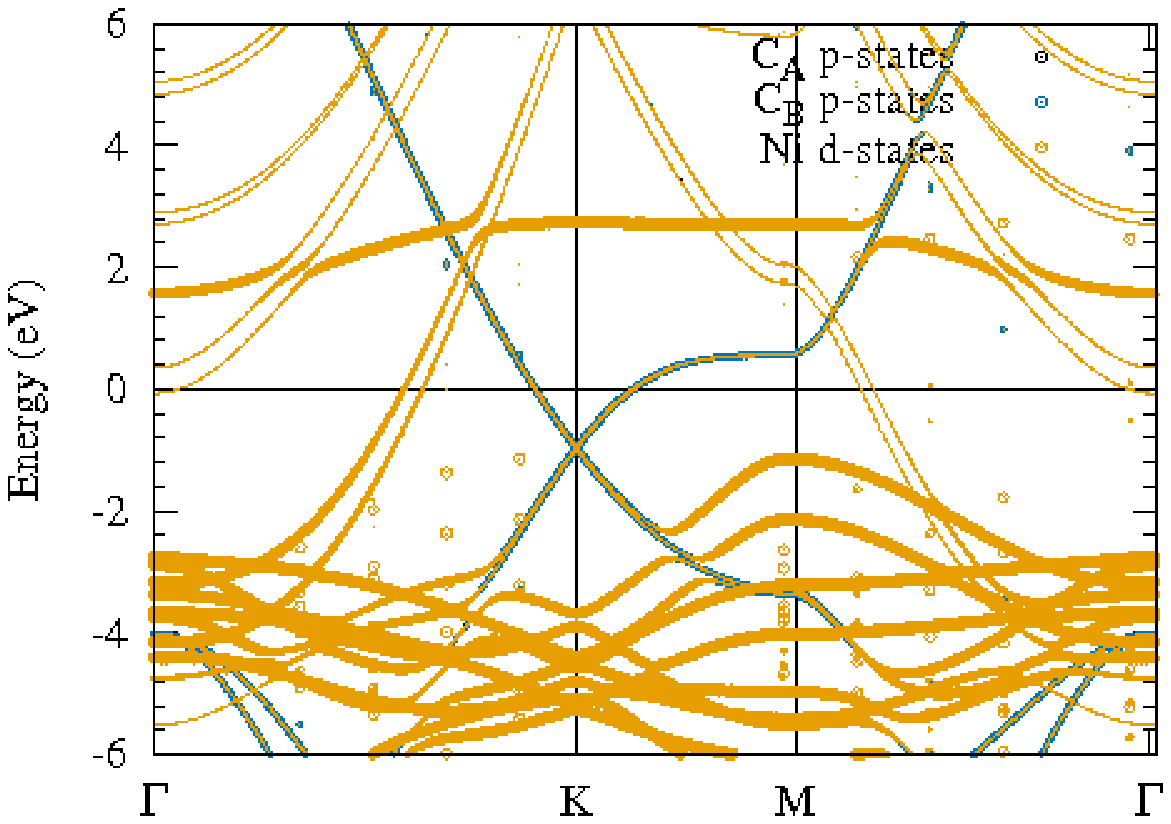}\\
\includegraphics[scale=0.45]{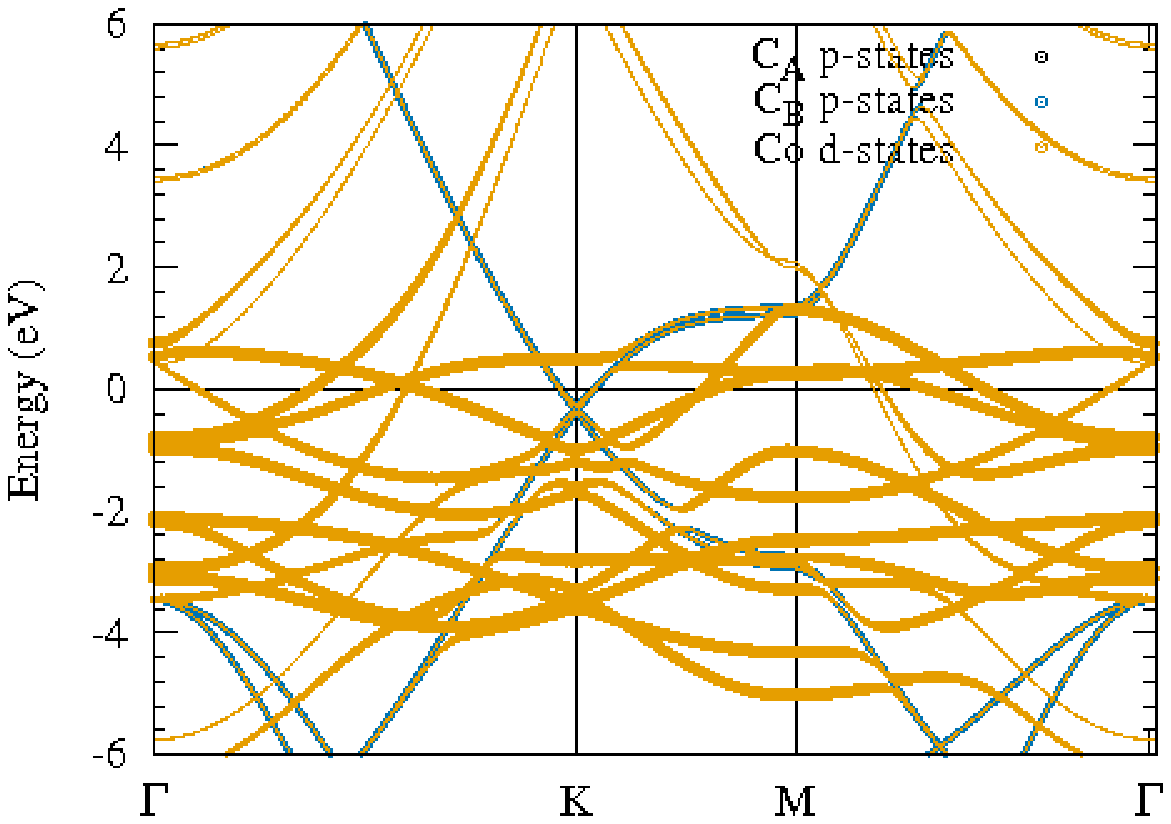}\\
\includegraphics[scale=0.45]{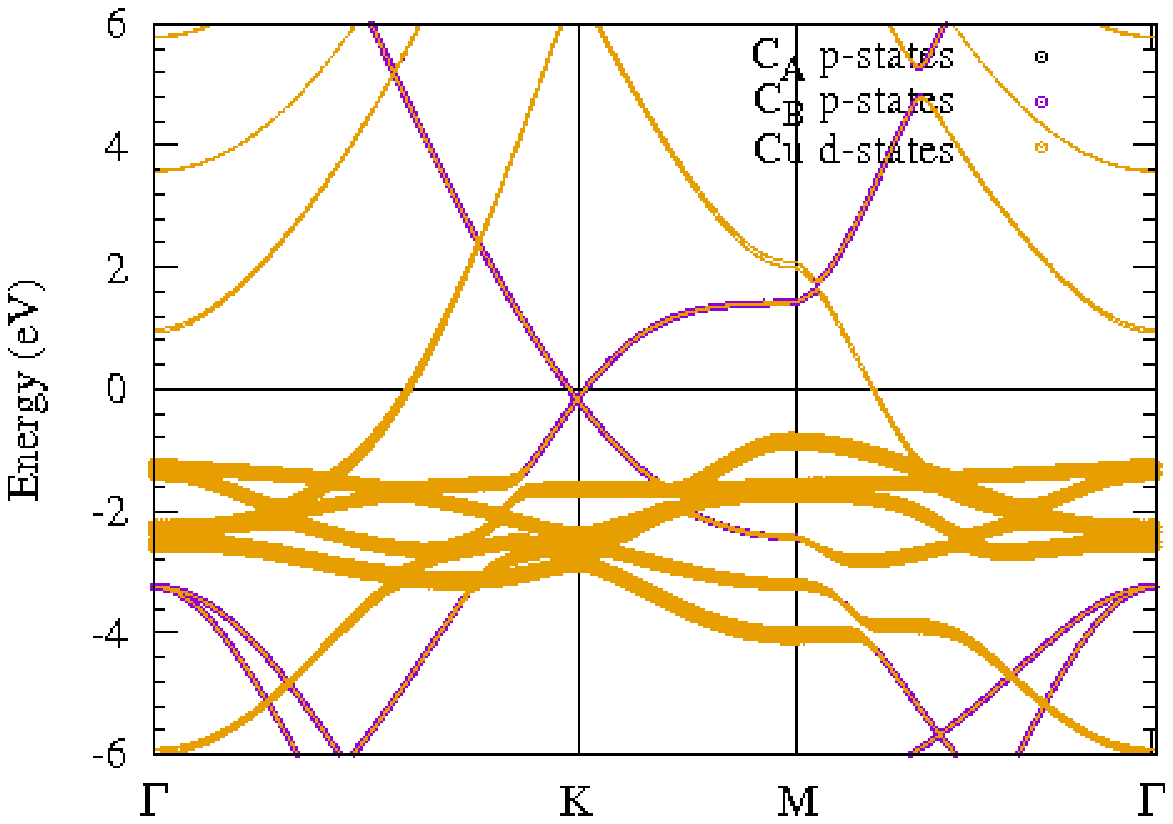}\\
\includegraphics[scale=0.45]{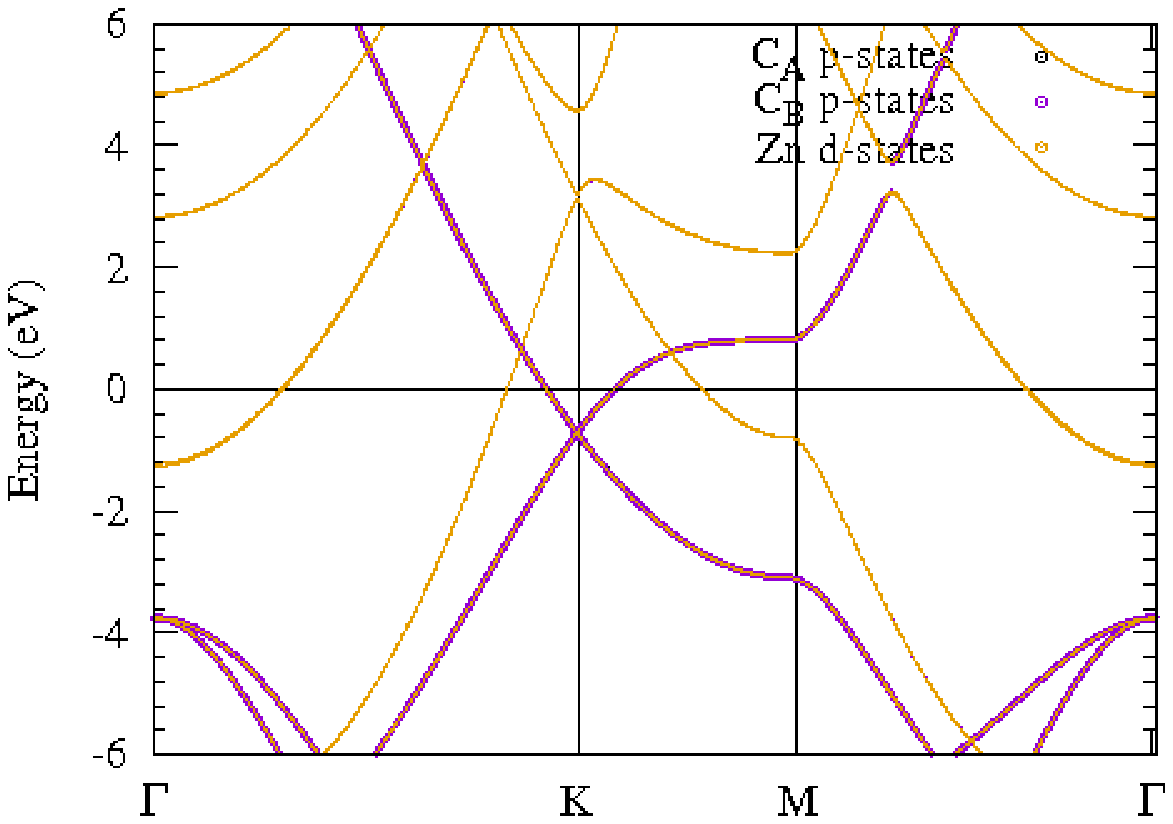}\\
\end{multicols}
\caption{\label{fatbands-h} The band structures of the graphene densely decorated 
with the $3d$ metallic adatoms and projected on the indicated type of atoms
(left panel). The middle panel presents the projection on adatomic 3$s$ and
3$p$ orbitals, as well as graphene 2$s$ states. The right hand side panel presents
the projection on the adatom 3$d$ states and graphene 2$p$ states. In the presented 
case the adatoms are placed in the {\it hollow} positions above the graphne.
The Hubbard corrections are taken into account within the calculations.
}
\end{figure}

\begin{figure}
\begin{multicols}{3}

\includegraphics[scale=0.45]{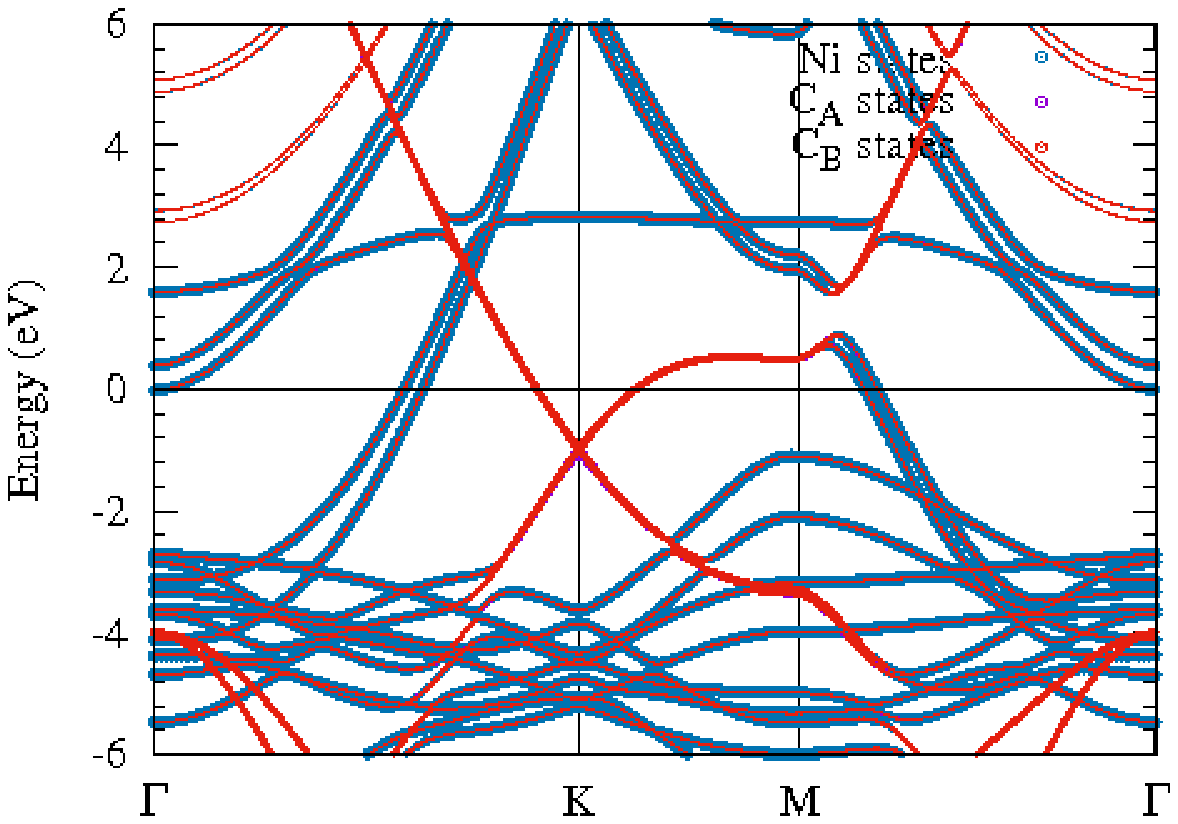}\\
\includegraphics[scale=0.45]{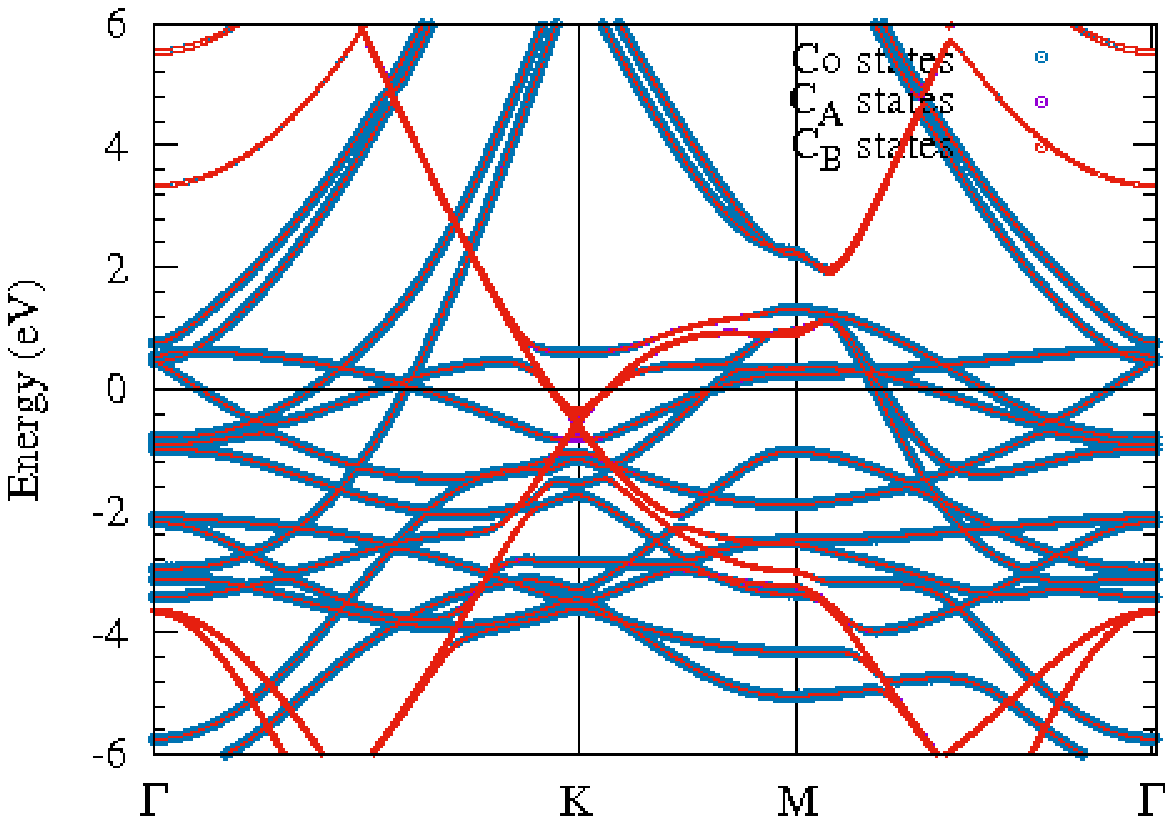}\\
\includegraphics[scale=0.45]{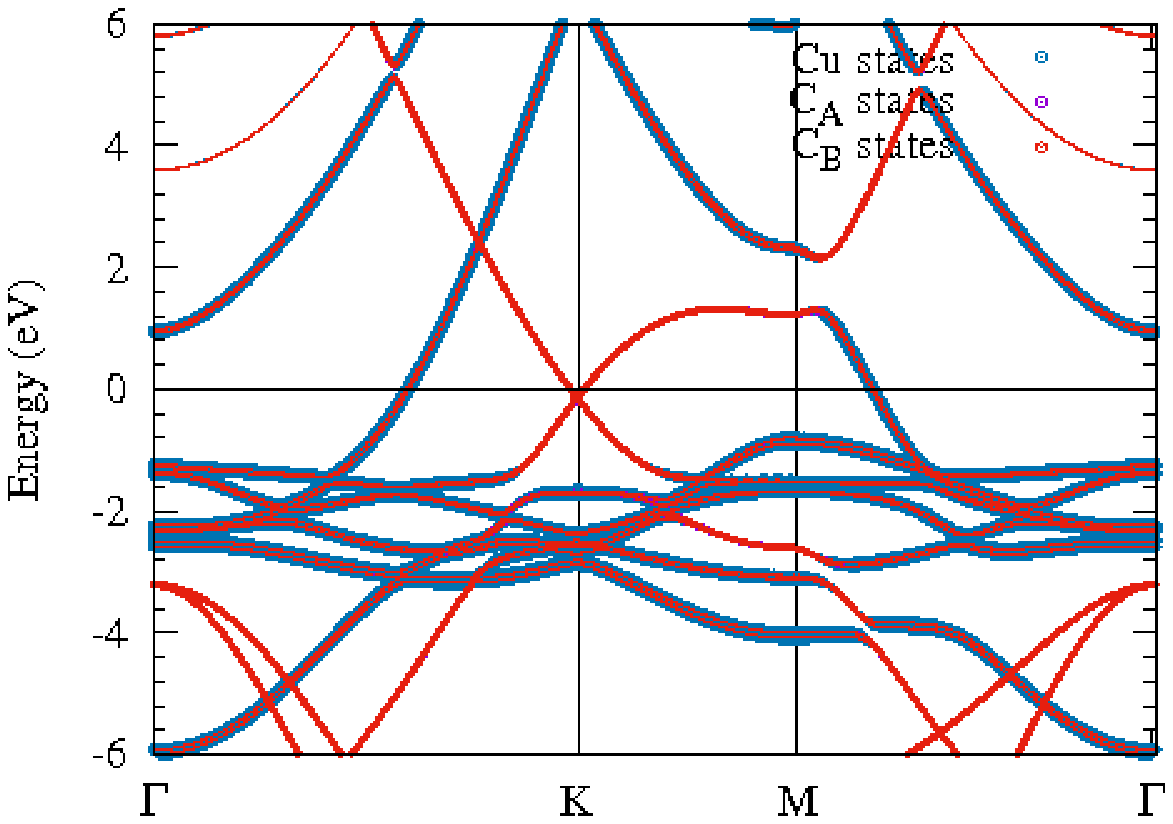}\\
\includegraphics[scale=0.45]{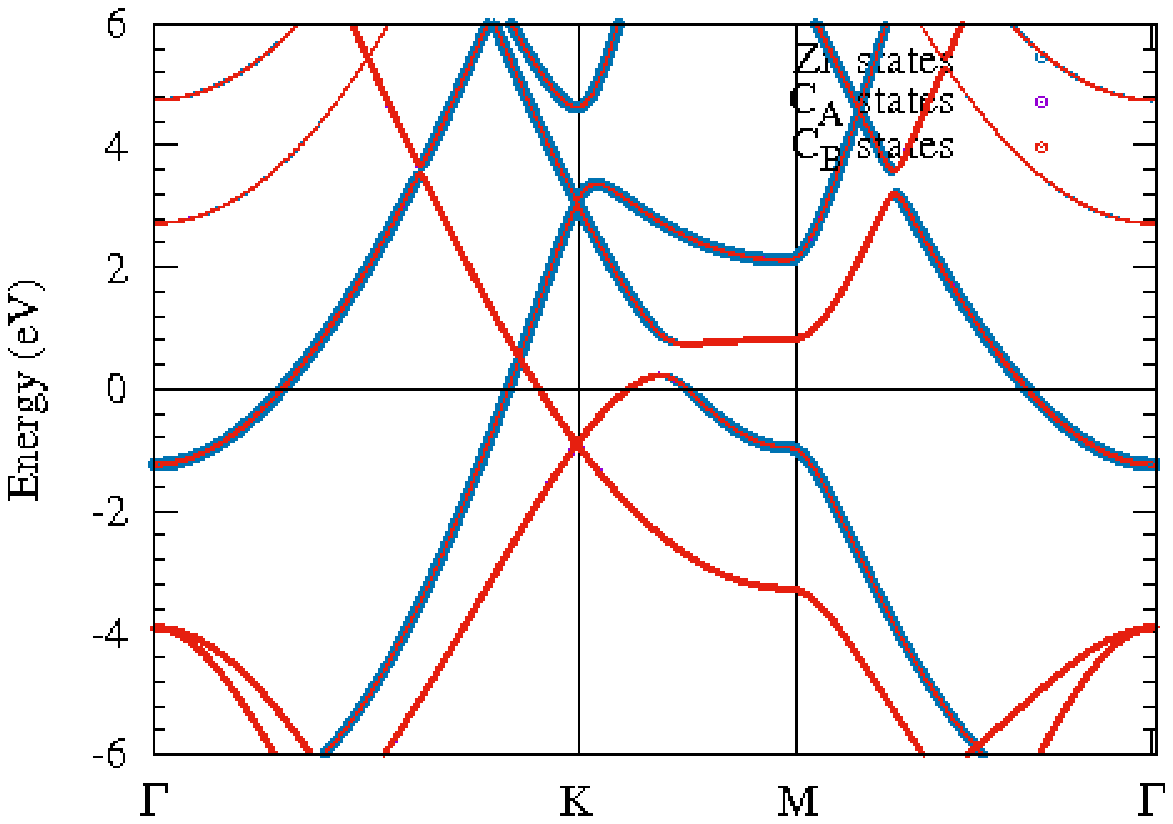}\\
\includegraphics[scale=0.45]{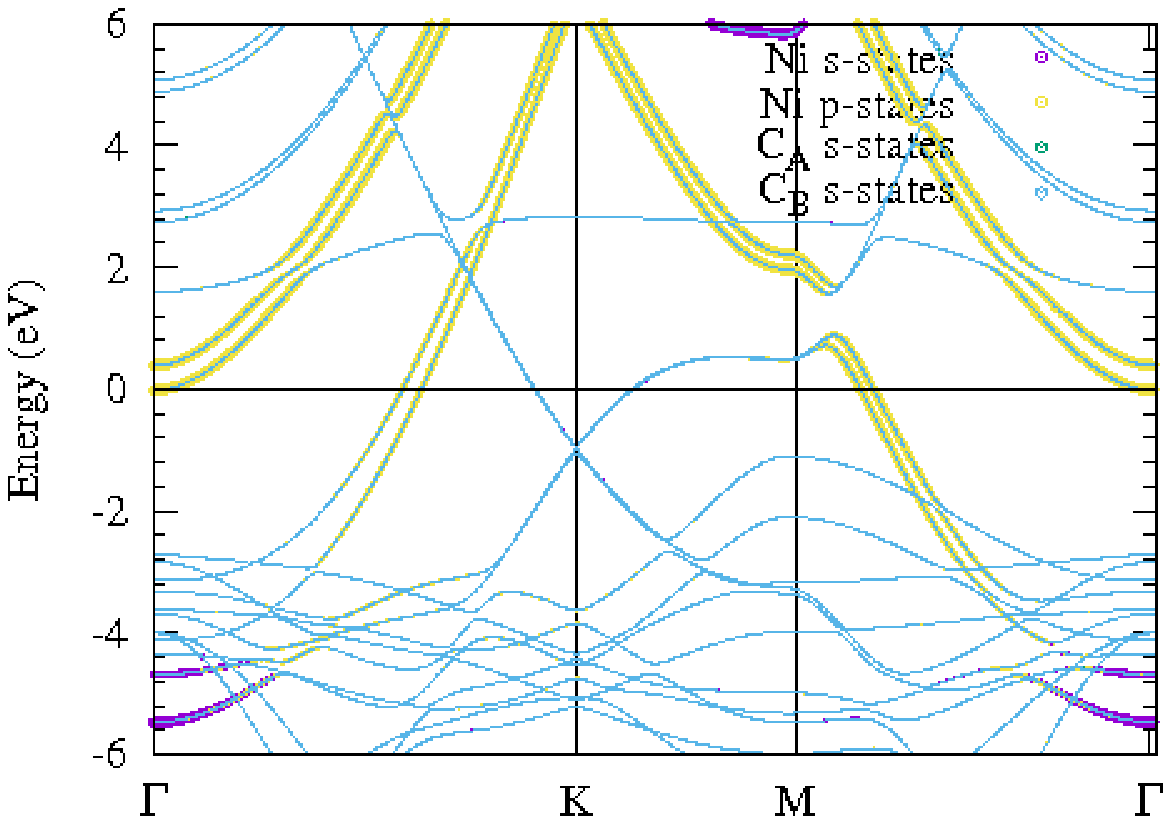}\\
\includegraphics[scale=0.45]{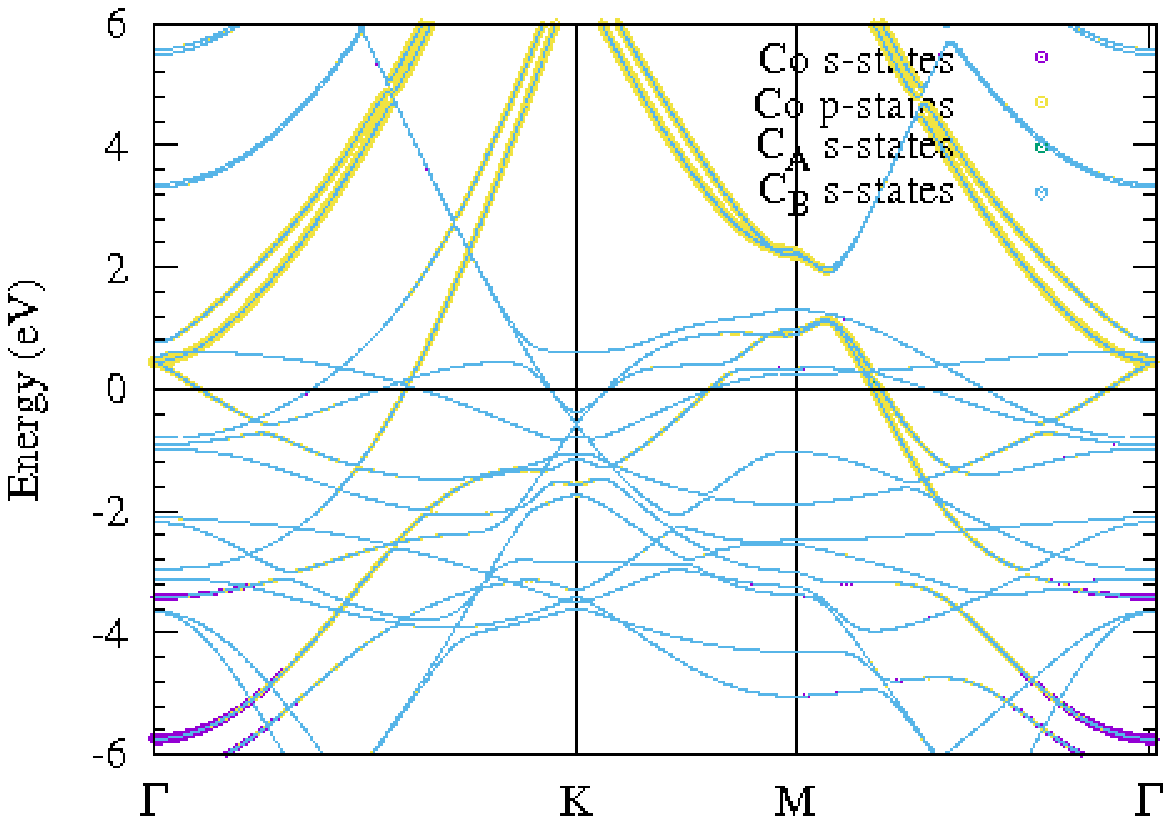}\\
\includegraphics[scale=0.45]{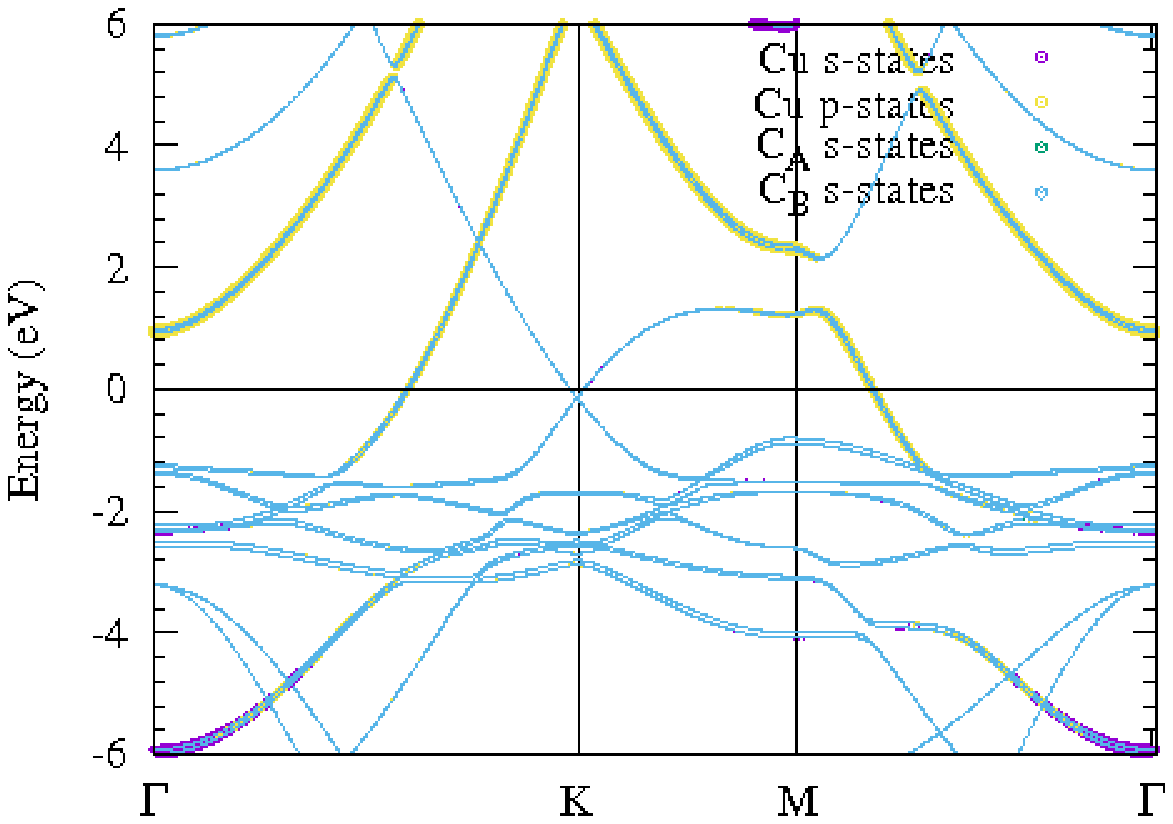}\\
\includegraphics[scale=0.45]{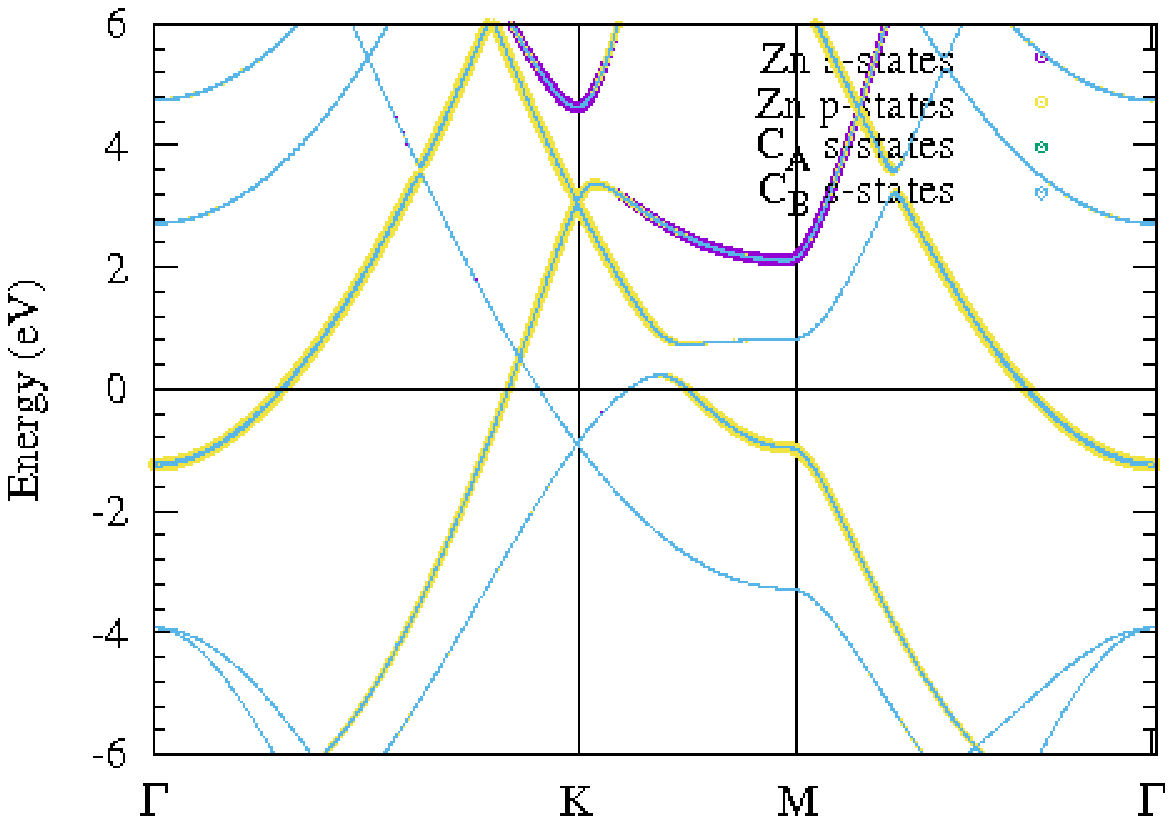}\\
\includegraphics[scale=0.45]{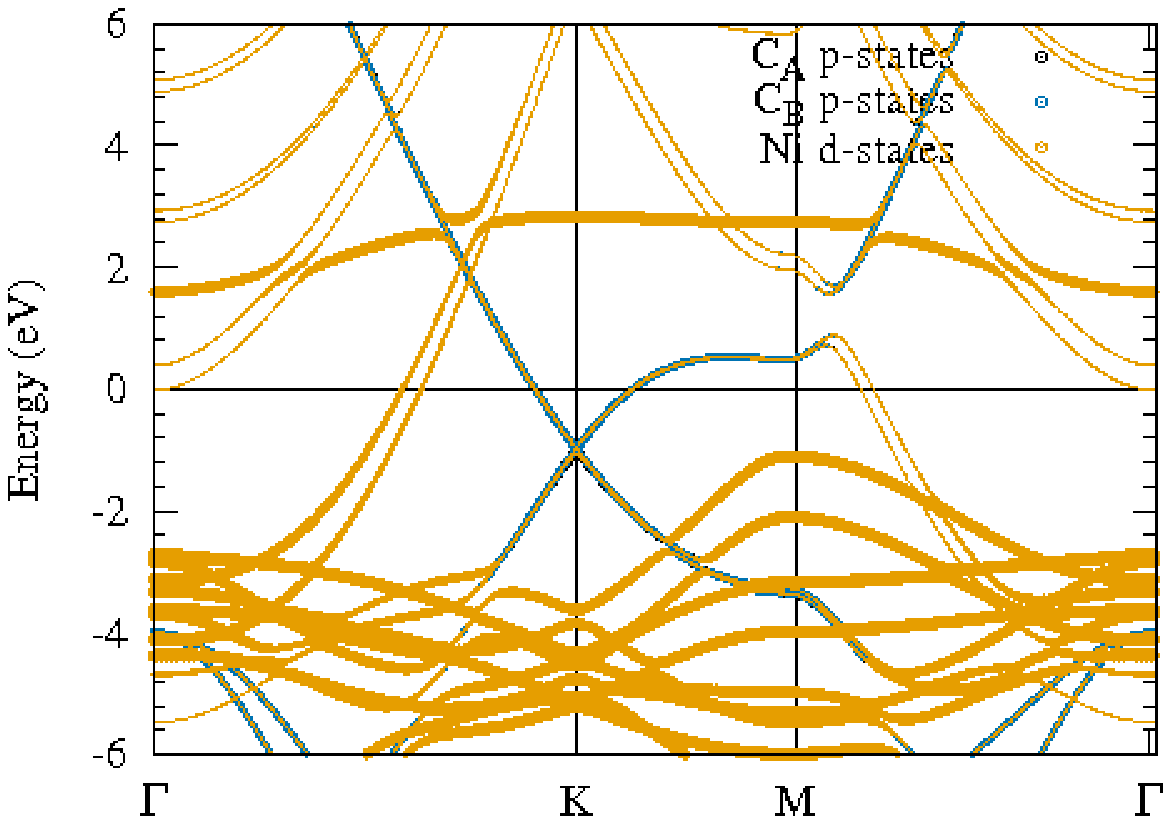}\\
\includegraphics[scale=0.45]{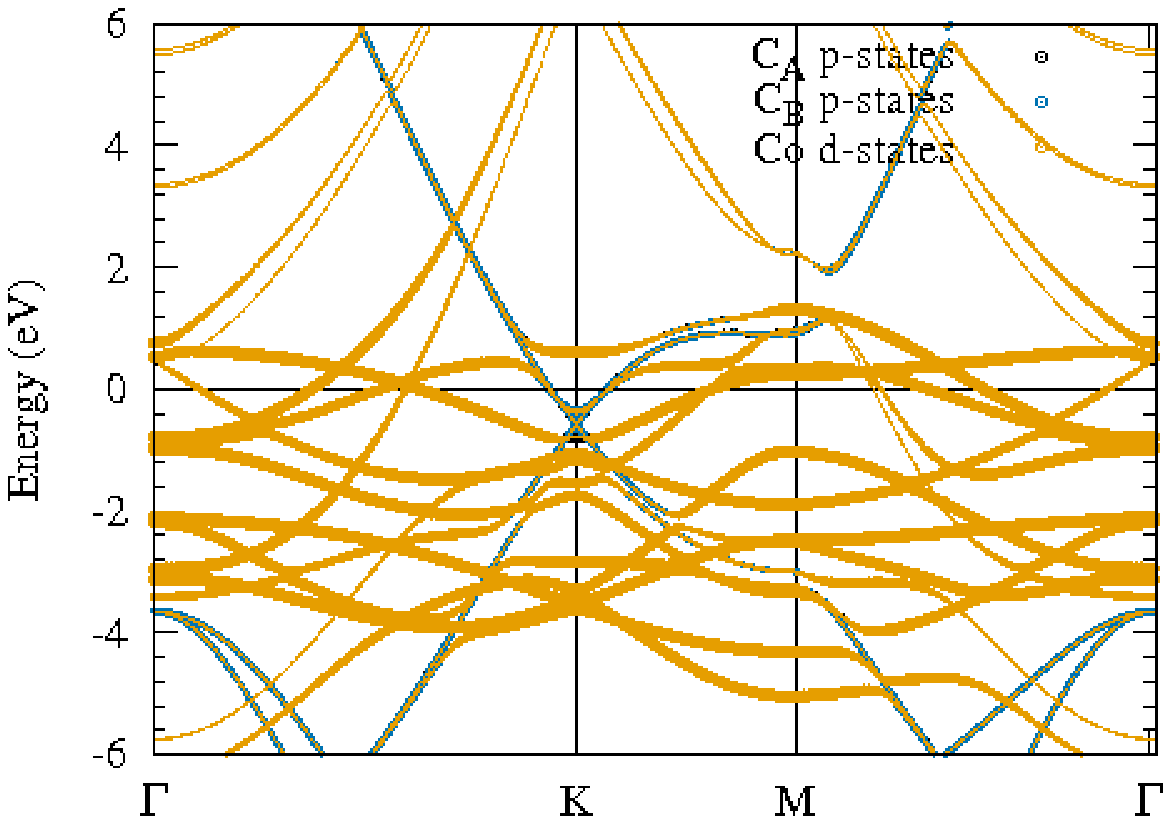}\\
\includegraphics[scale=0.45]{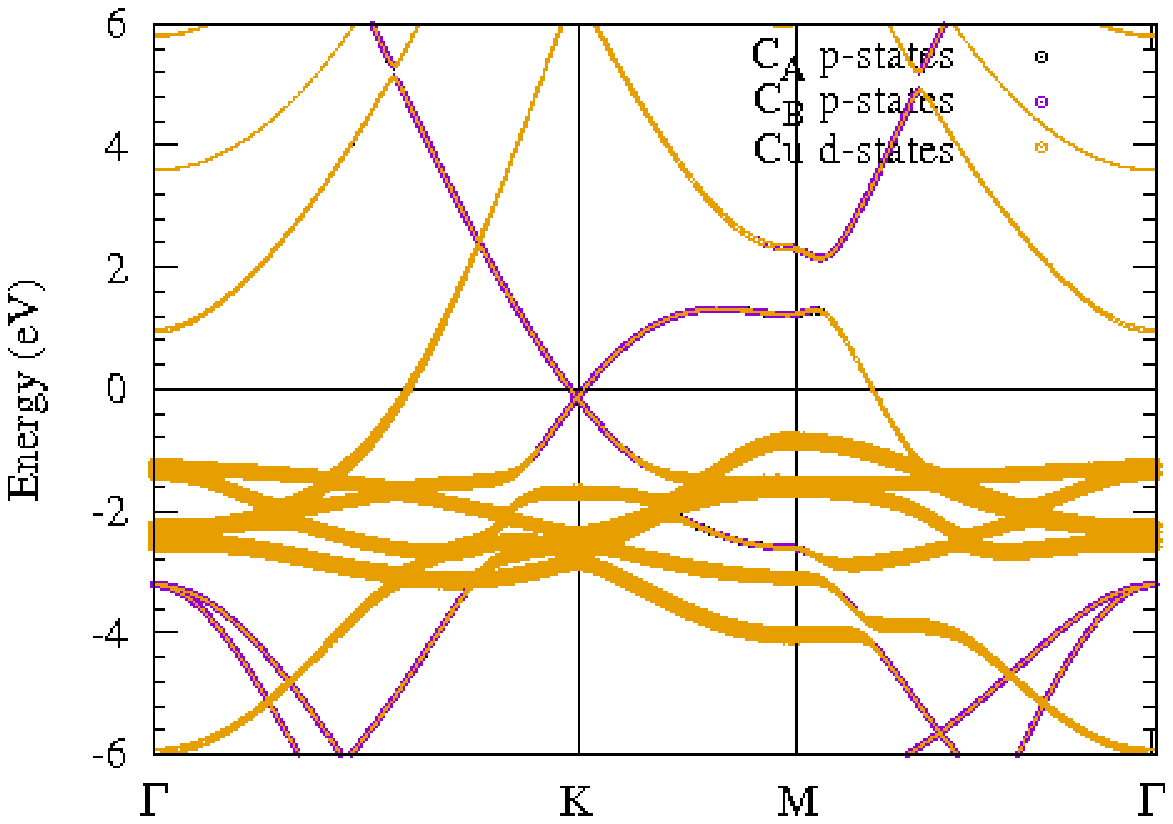}\\
\includegraphics[scale=0.45]{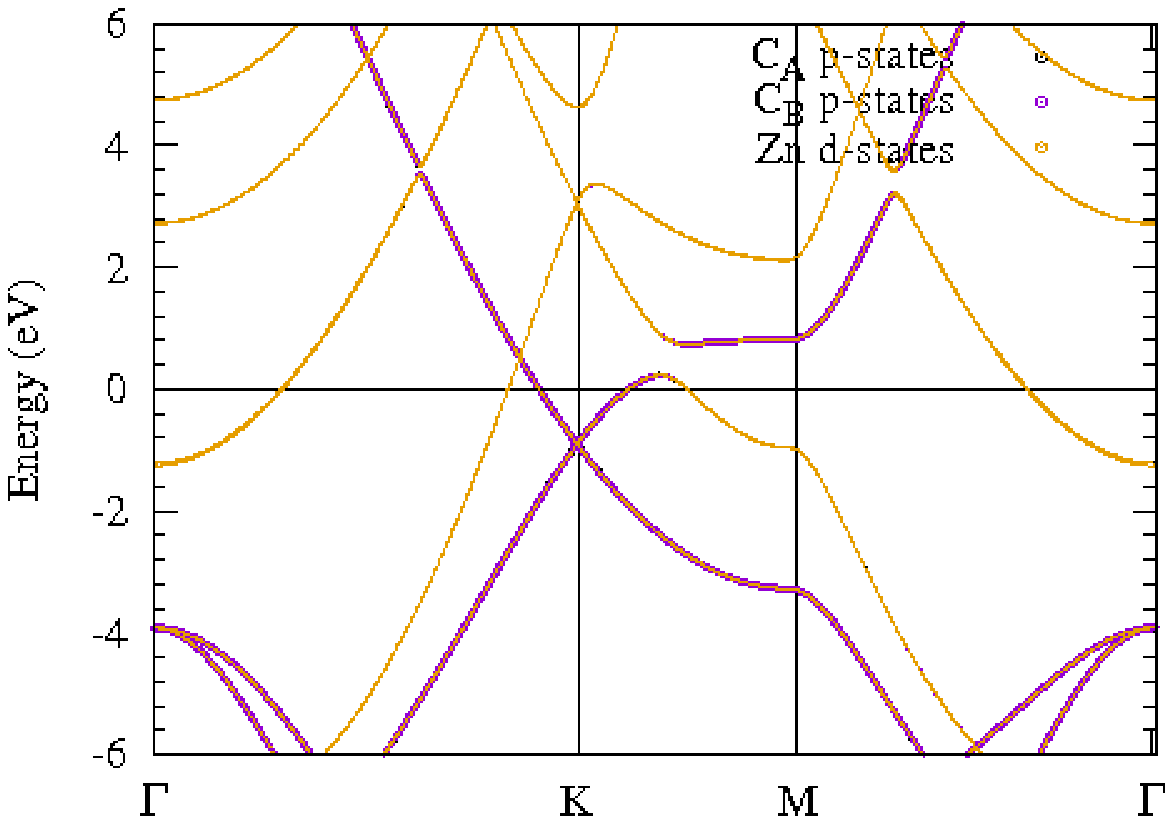}\\
\end{multicols}
\caption{\label{fatbands-t} The band structures of the graphene densely
decorated with the $3d$ metallic adatoms in the {\it on-top} positions above the
graphne, projected on the indicated atoms and states as in
Fig.\ref{fatbands-h}.}
\end{figure}

The analysis of the adatomic layer -- graphene hybridization for the {\it hollow}
and the {\it on-top} configurations is presented in Fig.\ref{fatbands-h} and
Fig.\ref{fatbands-t}, respectively. The left panel of these figures shows the band 
structure projected on the indicated atoms of the considered supercells. 
The provenance of the build-in flat doping bands, the modification of the 
graphene-based bands, as well as the crossings and anti-crossings between the adatomic 
and graphene states in the reciprocal space are visible. The middle panels of Fig.\ref{fatbands-h} 
and Fig.\ref{fatbands-t} show the band structures projected on the adatomic 3$p$ and 3$s$, 
as well as the graphene 2$s$ states, while the right panels show the projections on the 
adatomic 3$d$ and graphene 2$p$ states. 
In the case of the nickel, copper and zinc layers the grapehne 2$s$ and adatomic 3$s$, 3$p$ and 3$d$ 
states do not constitute the graphene cone. For the mentioned layers the graphene cone is constituted 
by the carbon 2$p$ states only and the contribution from the states of the A and B sublattices is equal
for the {\it hollow} configuration, while in the {\it on-top} configuration
the modest difference for the states of the A and B sublattives is observed. 
The exeption is the cobalt layer. For the cobalt in the {\it on-top}
confguration the $p-d$ hybridization encompasses the bands forming the Dirac
cone.
The flat doping bands originate from the adatomic 3$d$ states. For the cobalt layer these 
states are also found at the energy range where the Dirac point is located,
and for the conduction states. The hilly bands crossing the Fermi energy, that originate 
from the adatomic 3$p$ states for zinc, and from adatomic 3$p$ and 3$d$ states for nickel, cobalt and
carbon atoms, are also visible. Hence, the in-plane electron transport across the considered 
hybrid structure involves not only the graphene $\pi$-states, or in the case of the cobalt layer -- 
the low velocity doping states, but also the 3$p$ states for zinc layer and 3$p$ plus 3$d$ states 
of the nickel, cobalt and copper metallic layers. This analysis provides us the information for
the proper construction of an effective Hamiltonian in the charge carrier transport description, where 
not only the graphene 2$p$ and adatomic (or substrate) 3$d$ states should be taken into account, but 
also the 3$p$ doping states should be included.

\subsection{The magnetism induced in graphene due to the proximity of an adatomic metallic layer}

Due to the proximity of the adatomic layers the finite values of the magnetic moments are induced on 
the graphene sublattices. The induced magnetism and its impact on the band structure depends significantly 
on the adatom type (whether they are magnetic or not) and the configuration on the graphene lattice. 

Although the total magnetic moment of the considered supercells containing nonmagnetic adatoms is
zero, we find the small values of the induced magnetic moments on the graphene as well as on the 
metallic adatoms, for the studied nonmagnetic metallic layers.
Hence, the finite values of the exchange coupling should be taken into account while describing 
the system with an effective Hamiltonian. The analysis of the proposed
Hamiltonian is discussed in the next part of this paper. We also notice, that the tiny induced magnetic 
moments on the adatoms and the graphene sublattices are oriented perpendicularly to the system surface 
and oriented in line.

For the ferromagnetic layers we observe that for the {\it on-top}, as well as for the {\it hollow} 
configurations, magnetic moments located on the adatoms and those induced on graphene, are oriented perpendicularly 
to the surface of the two-layered structure. However, the aligment of the magnetic moments differs for {\it on-top} 
and {\it hollow} configurations. 
In the {\it on-top} configuration, the alignment of the magnetic moments of the carbon atoms located 
below the adatom (let us assume it is the graphene A sublattice) is opposite to the aligment of the 
magnetic moments on the adatoms, while the magnetic moments induced on the other graphene sublattice -- B,  
are aligned in the same direction as the moments on the adatoms. 
In the {\it hollow} configuration the aligment of the induced magnetic moments on the both graphene sublattices 
is opposite (cobalt) or consistent (nickel) to the direction of the magnetic moments localized on the adatoms. 
Hence, in the {\it on-top} configuration, the proximity of the adatomic layer causes the anti-ferromagnetic aliment on 
the graphene sublattices, while in the {\it hollow} configuration the ferromagnetic alignment on the graphene
sublattices is forced by the proximity of the localized magnetic moment, that equally couples to both graphene
sublattices. 
 
The values of the magnetic moments located on the specified magnetic atoms for the studied structures 
are presented in Tab.\ref{mag}. The proximity of all studied adatomic layer causes the
exchange splitting of the graphene bands. This splitting is pronouced for the ferromagnetic adatomic 
layers and influences the electronic properties of the system. 
The exchange coupling and resulting magnetic splitting effect is particularly important 
in analyzing the band structure landscape near the Dirac point.
 
\begin{table}
\caption{The values of magnetic moments on the specified adatoms and induced
on the carbon atoms of the graphene A and B sublattices, in the $\mu_{\rm
B}$, for the {\it hollow} and {\it on-top} configurations. In the case of the 
{\it on-top} configuration the adatoms are located above the A sublatice. 

\label{mag}}
\begin{center}
  \begin{tabular}{  l  c  c  c  c  c  c }
    \hline \hline
        &  Ni       & C$_{\rm A}$ (Ni) &   C$_{\rm B}$ (Ni) & Co     & C$_{\rm A}$(Co) & C$_{\rm B}$(Co)  \\ \hline
 on-top & 1.0521    & -0.0031          &  0.0012            & 1.8458 & -0.0069         &  0.0067    \\ 
 hollow & 1.0503    &  0.0045          &  0.0045            & 1.9232 & -0.0008         & -0.0008    \\
    \hline \hline
  \end{tabular}
\end{center}
\end{table}

\subsection{Energy dispersion in the vicinity of the Dirac point}

In the case of the proximity of the metallic layer to the graphene, the electronic properties near 
the Dirac point can be mapped on the following effective Hamiltonian \cite{Zollner,Dyrdal2D}
\begin{equation}
\label{hamTOT}
H=H_0+H_{\Delta}+H_{\rm exch}+H_R,
\end{equation}
where $H_0=\hbar v_{\rm D} (\tau k_x\sigma_x + k_y\sigma_y)$ describes the
kinetic energy of the charge carriers, with $k_x$ and $k_y$ dentoting the
components of the charge carrier wave function, and $\sigma_x$, $\sigma_y$ denote
the Pauli matrices, while $v_{\rm D}$ stands for the electron velocity at
the Dirac point and $\tau=\pm 1$ allows to distinguish between the $K(K^{'})$ 
points. This term refers to the gapless Dirac states near $K(K^{'})$ points. 
The proximity of the metallic layer to the graphene results in symmetry 
beaking of the pseudospins attributed to the graphene sublattices and the 
formation of the staggered potential that is felt by the pseudospins. 
This symmetry beaking results in formation an energy gap of the width 2$\Delta$ 
at the Dirac point, and can be described with the Hamiltonian $H_{\Delta}=\Delta\sigma_z s_{0}$, 
where the $s_0$, denotes the unit matrix in the spin-space, while $\sigma_z$ denotes
the Pauli matrix. The proximity exchange effects are then described by the Hamiltonian \cite{Zollner} 
\begin{equation}
\label{hamEXCH}
H_{\rm exch}=\lambda_{A}[(\sigma_z+\sigma_0)/2]s_z+\lambda_{B}[(\sigma_z - \sigma_0)/2]s_z, 
\end{equation}
with the $\lambda_A$ and $\lambda_B$ being the exchange coupling constants (in the sense of the magnetic 
exchange coupling constants) for the indicated graphene sublattice. We notice, that in the {\it hollow} 
configuration, for the adatoms equally coupled to both graphene sublattices  $\lambda_A \approx
-\lambda_B$. 
For the {\it on-top} configuration, when the coupling to one of the graphene sublattices (let us assume A) 
is much stronger in comparsion to the coupling with the other subllatice, $\lambda_A >> \lambda_B$.

Finaly, the proximity effect may lead to the enhancement of the Rashba-like interaction on the graphene, 
hence one should also take into account the term \cite{Dyrdal2D} 
\begin{equation}
\label{hamR}
H_R= \lambda_R(\tau\sigma_x s_y -\sigma_y s_x),
\end{equation} 
The mapping of the described effective Hamiltonian on the DFT analysis, given in allows us to extract 
the parameters, that are gathered in Tab.\ref{param}.

\begin{table}
\caption{The values of indicated parameters of the Hamiltonian given by Eq.\ref{hamTOT}
in meV, for the studied adatomic layers in the {\it hollow} (h) and {\it on-top} (t) configurations.
\label{param}}
\begin{center}
  \begin{tabular}{  l  c  r  r  r }
    \hline \hline
                 & $\Delta$      &   $\lambda_A$             &  $\lambda_B$       & $\lambda_R$       \\ \hline
Cu(t)            &  7.885        &   1.973                   &  0.159             &  1.6               \\
Cu(h)            &  --           &   1.829                   & -1.849             &  1.8               \\
Zn(t)            & 19.663        &   1.726                   &  1.367             &  1.5               \\
Zn(h)            &  --           &   0.228                   & -0.208             &  0.2               \\
Ni(t)            & 29.534        &  60.820                   &  8.907             &  4.9               \\  
Ni(h)            &  --           & -49.287                   & 50.460             &  4.6               \\
Co(t)            & 53.751        &  72.654                   &  4.732             &  5.3         \\
Co(h)            &  --           &  67.853                   &-71.224             &  5.0         \\        
    \hline \hline
  \end{tabular}
\end{center}
\end{table}

\begin{figure}
\begin{multicols}{2}
\label{CuZn-cone}
\includegraphics[scale=0.5]{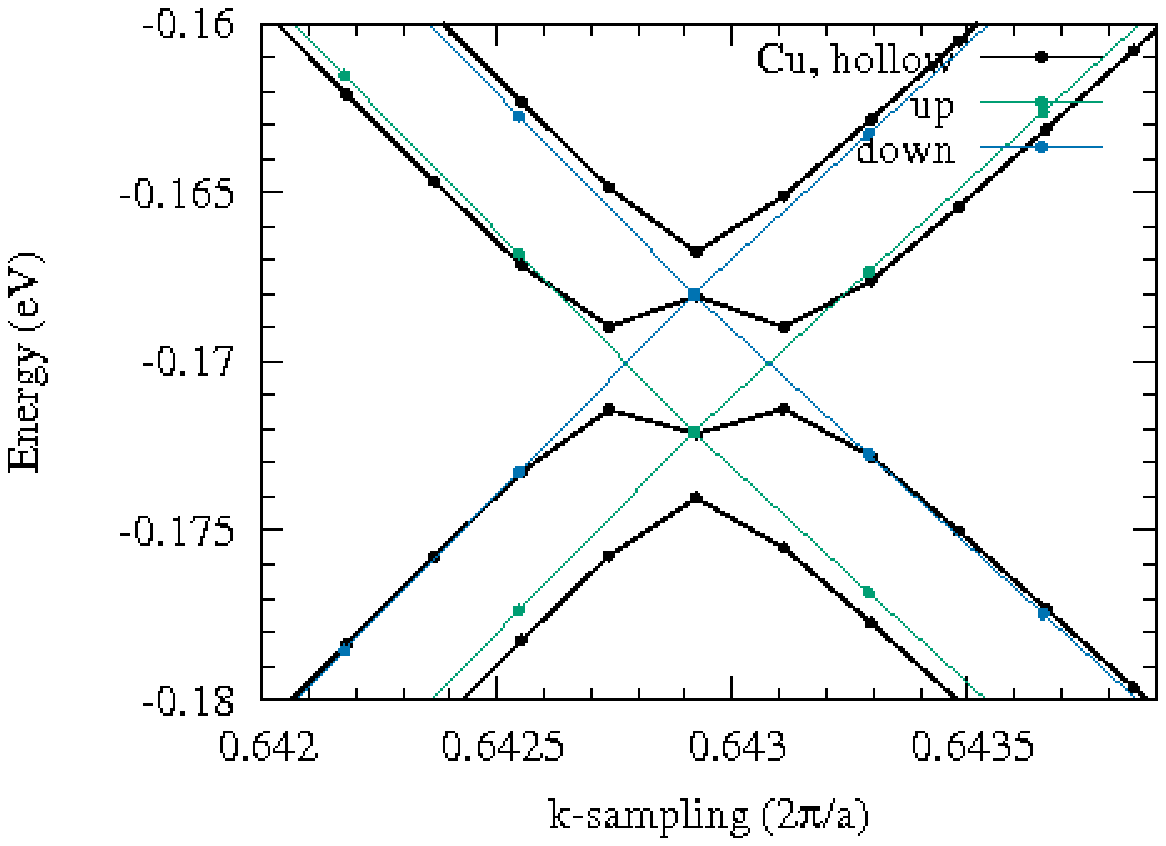}\\
\includegraphics[scale=0.5]{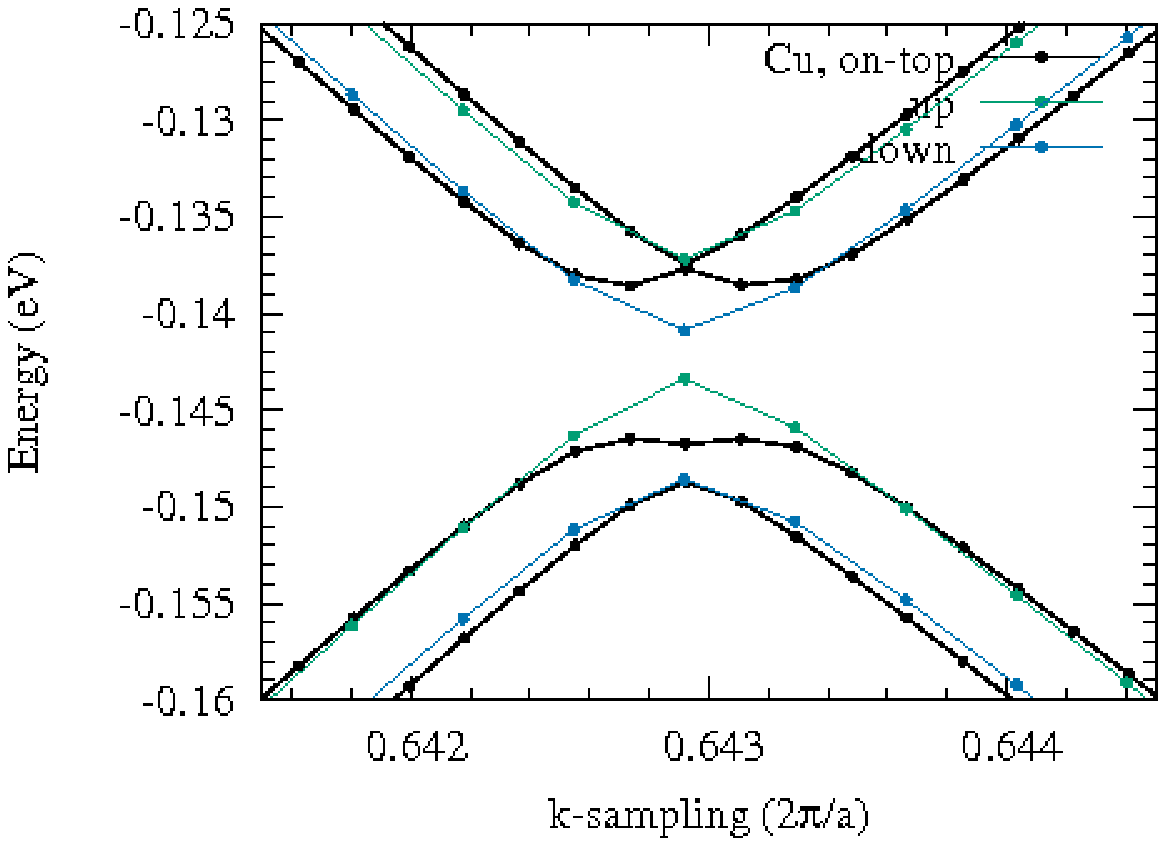}\\
\includegraphics[scale=0.5]{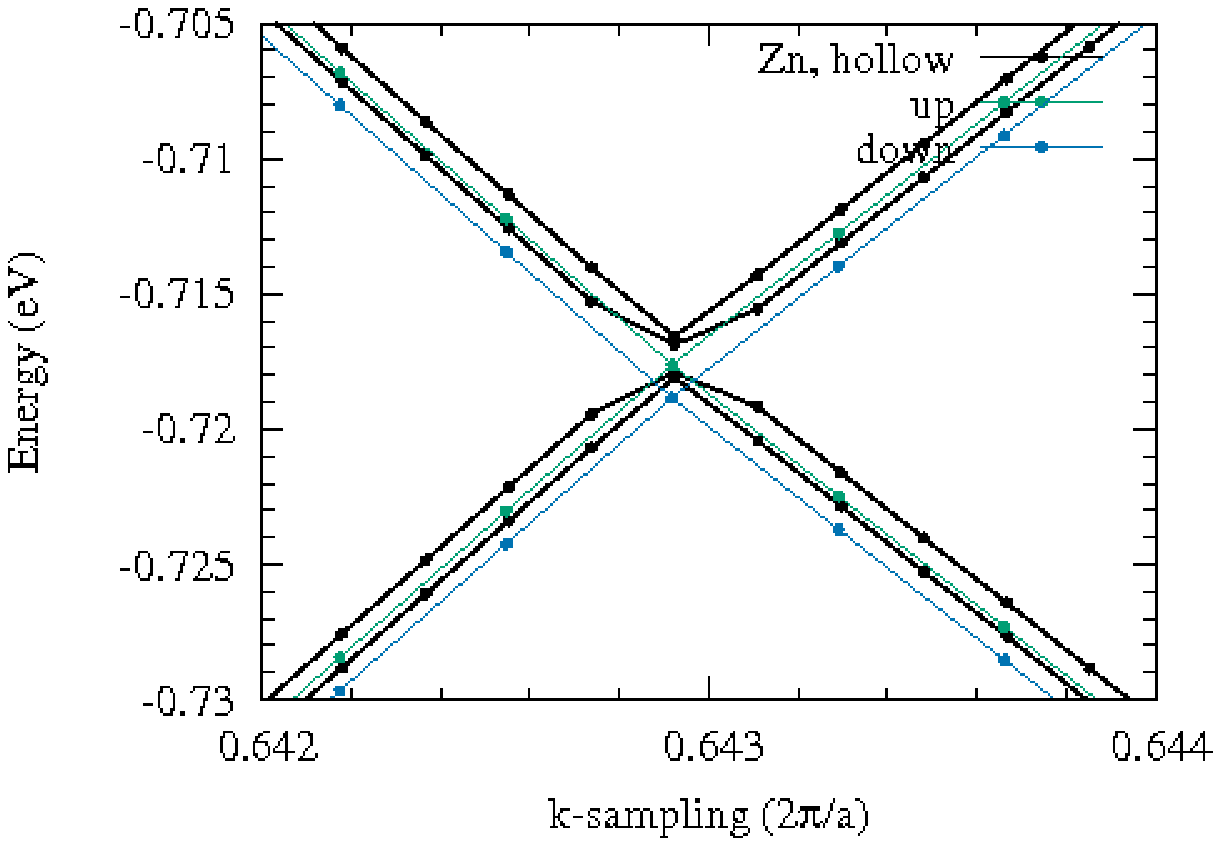}\\
\includegraphics[scale=0.5]{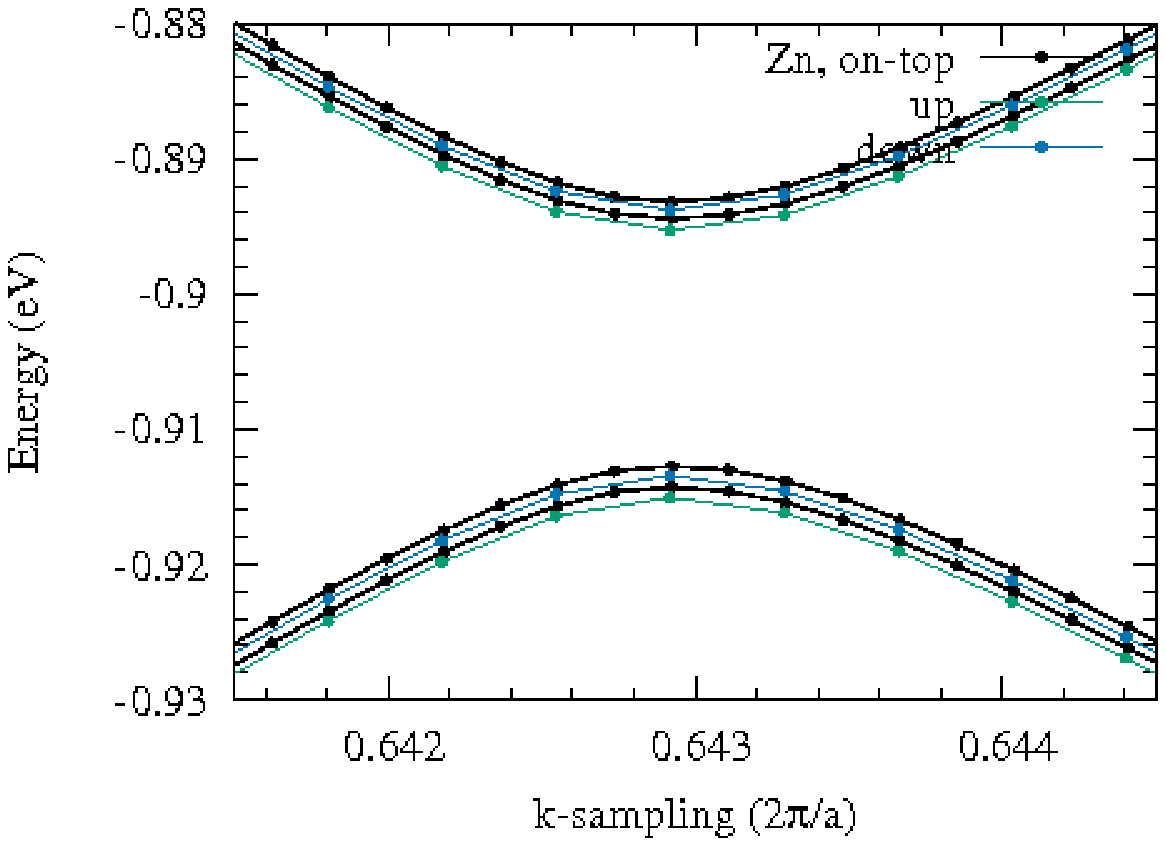}\\
\includegraphics[scale=0.5]{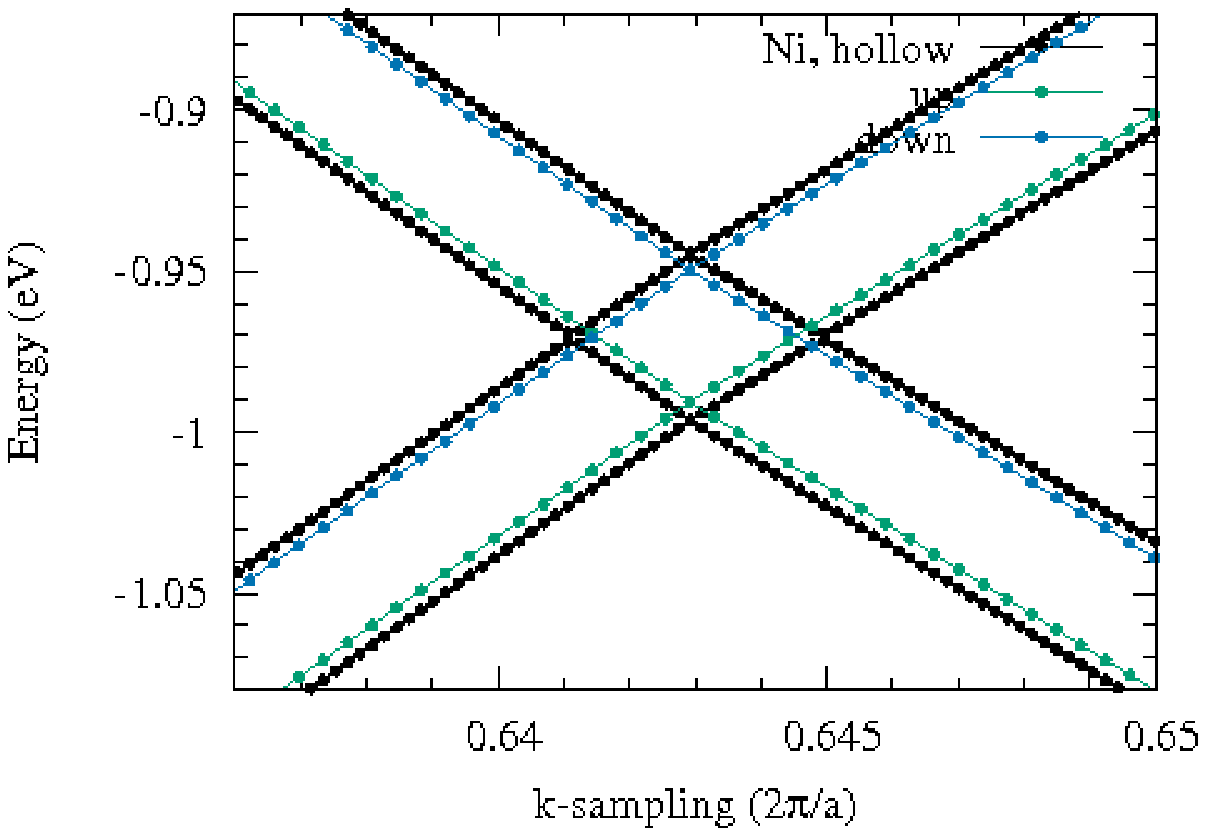}\\
\includegraphics[scale=0.5]{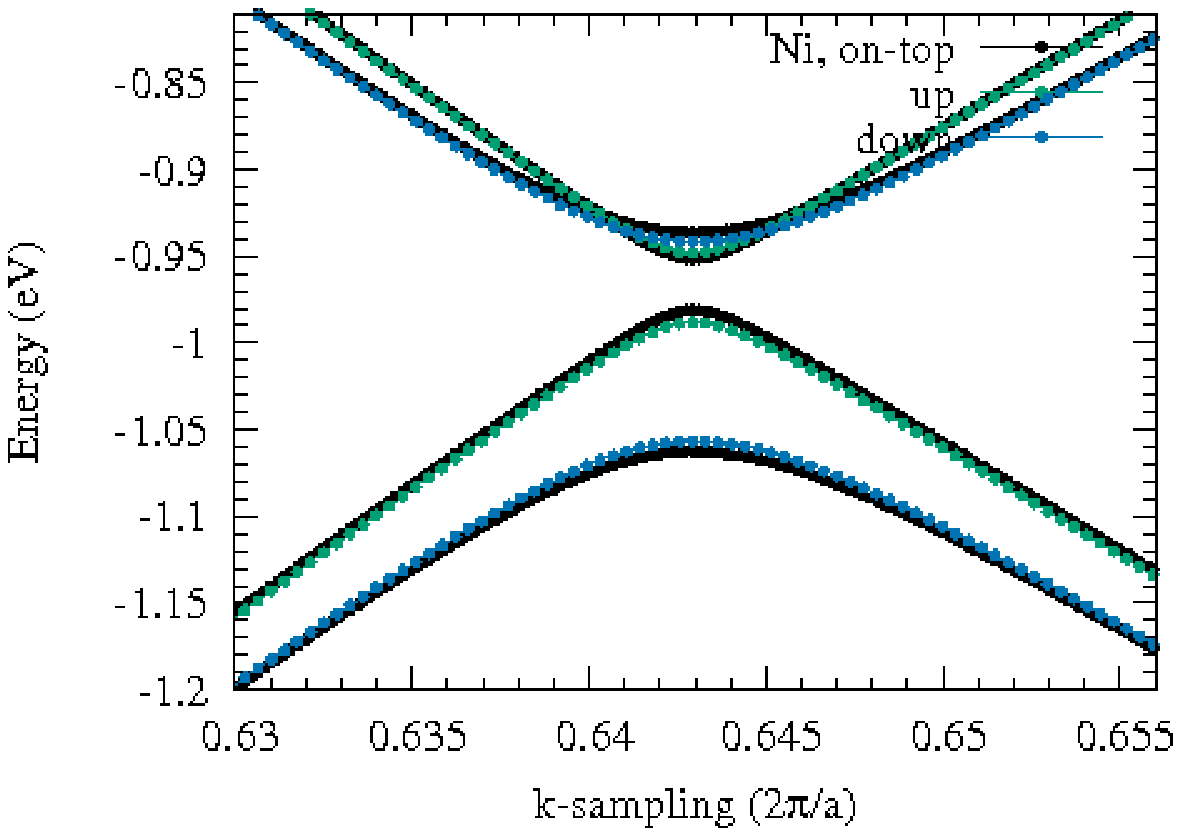}\\
\includegraphics[scale=0.5]{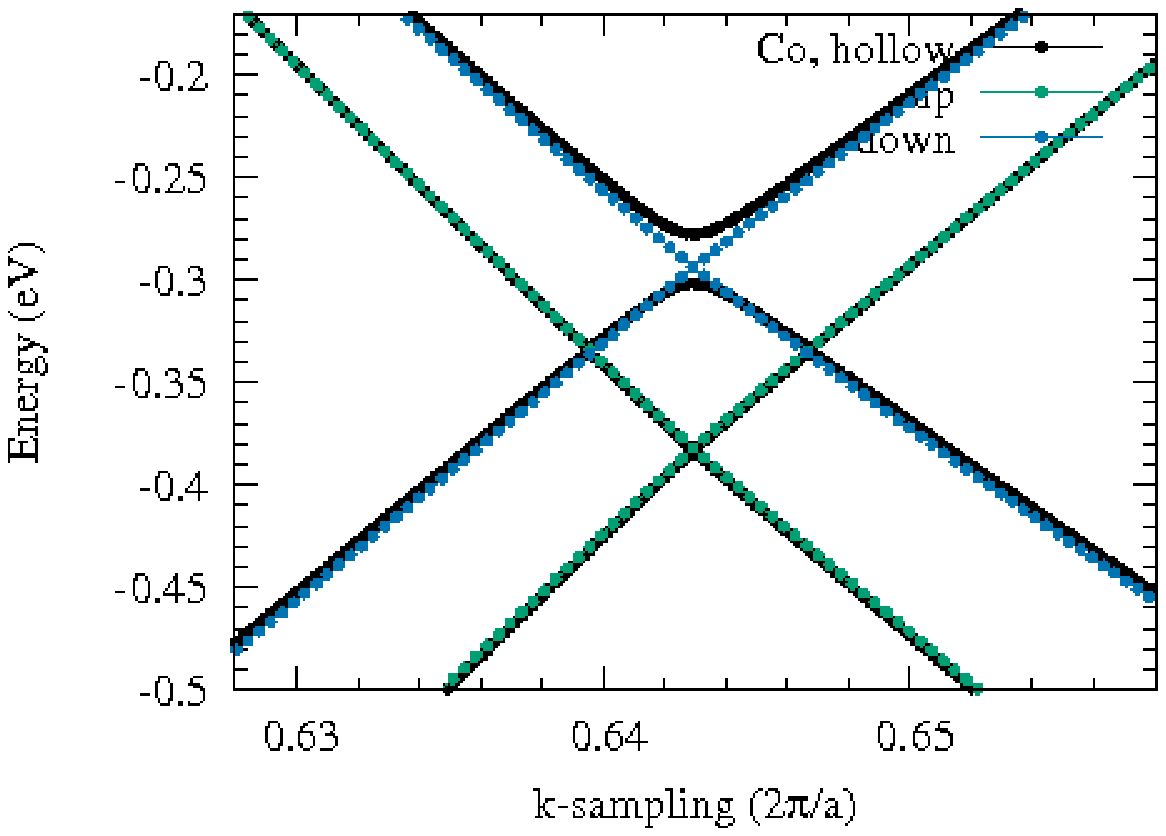}\\
\includegraphics[scale=0.5]{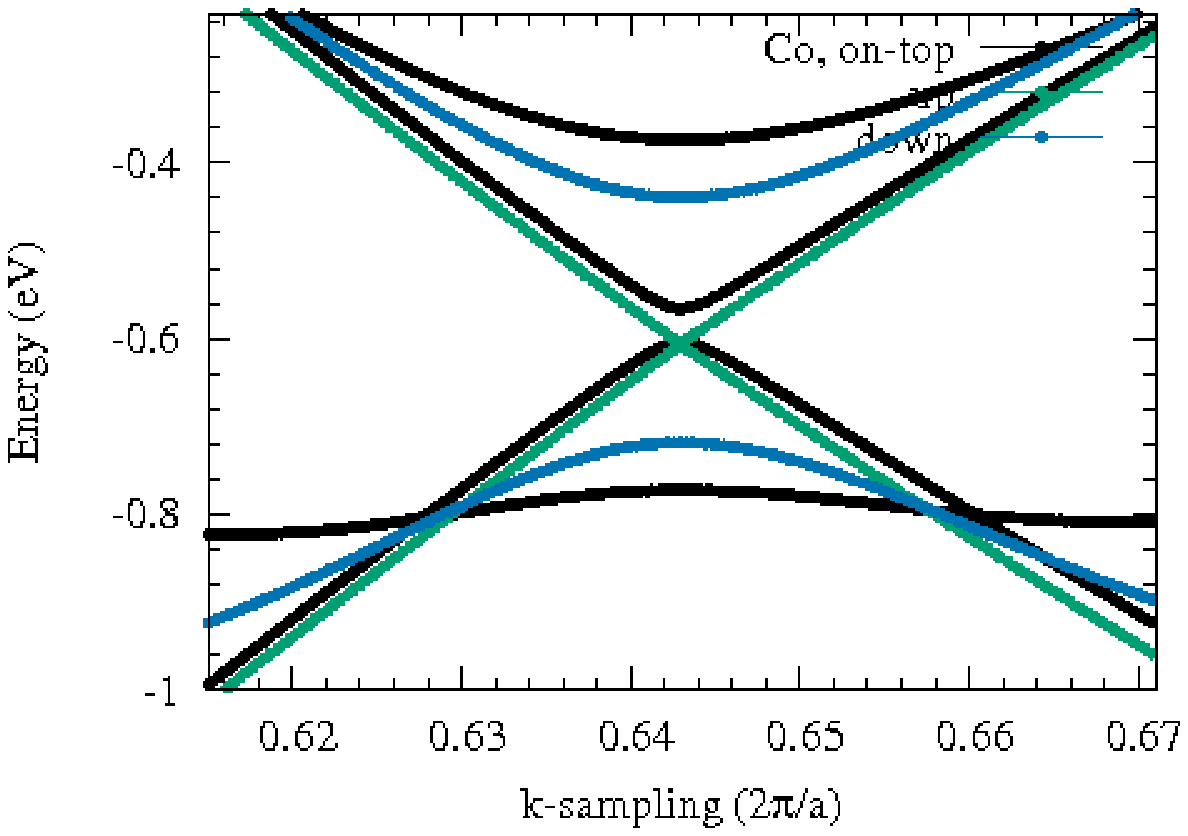}
\end{multicols}
\caption{The band structure in the vicinity of the Dirac point for the {\it
on-top} and {\it hollow} configurations for all considered adatomic metallic
layers. The band structure for the spin-polarized case, when no spin-orbit
coupling is taken in to account is also shown for compartion.}
\end{figure}

The band structures near the Dirac point for the graphene in the proximity of the copper 
and zinc metallic layers are presented in Fig.\ref{CuZn-cone}.
The general feature can be seen, namely, due to the staggered potential in the
{\it on-top} configuration the energy gap at the $K(K^{'})$ is formed for all studied
adatomic layers and this energy gap is additionaly widen by the Rashba-like
interaction. On the other hand, for the {\it hollow} configuration the energy gap at the
$K(K^{'})$ results from the spin-orbit coupling only. This is clearly seen
when comparing the band structures at the $K$ point for the case when the
spin-orbit interaction is included in the DFT calculation, with the case
when calculations are performed assuming only spin-polarized system.  


\section{Conclusions}
Within this work we studied the electronic and magnetic properties of the two-layered structure, 
namely when the graphene is decorated densely with the $3d$ metallic
adatoms. We found that no covalent bonds are formed between the graphene carbon atoms and
the metallic adatoms for the perfect dense decoration. The $n$-type doping of the graphene 
is found for all studied cases and is caused by the charge transfer from the metallic layer to the
graphene. This charge transfer involves the orbital hybridization of the $2p$
graphene states and $3d$ as well as $3p$ states of the adatomic layer. The
two types of the adatomic layers may be distinguished among the studied
layers, namely the adatoms that interact strongly with the graphene (nickel,
cobalt and zinc) and those which interact weakly -- copper.
The proximity of the metallic layer leads to the modification of the
graphene electronic properties not only {\it via} the graphene charge doping but 
also through the exchange coupling arising between the graphene and metallic
layers, the shape of the staggered potential and the spin-orbit Rashba-like interaction.
The exchange coupling is pronauced for the ferromagnetic layers, however we
find the small finite values of the exchange coupling also for the nonmagnetic
layers. 
We show that the proximity effect visibly modifies the Dirac cone and this
modification depends on the adatomic configuration, namely if the adatom is
coupled equally to both graphene sublattices ({\it hollow}) or the coupling
with one of the sublattices dominates ({\it on-top}).  
In the case of the {\it on-top} as well as {\it hollow} configurations the band 
spin-splitting arises due to the exchange coupling of the graphene and the
metallic layer. This spin-splitting is of the the order of magnitude
larger for 
the ferromagnetic layers than for the non-magnetic layers. Additionally, for 
the {\it on-top} configuration the energy gap apprears, due to the staggered potential. 
This energy gap is then modified by the spin-orbit Rashba-like interaction,
although this modification is insignificant in comparsion to the value of the gap. 
On the other hand, for the {\it hollow} configuration, that Rashba-like interaction
introduces the energy gap at the Dirac point, since the contribution form
the staggered potential is absent. 

\section*{Acknowledgments}
This work has been supported by the Polish Ministry of Science and Higher Education through a research
project ’Iuventus Plus’ in years 2015-2017 (project No. 0083/IP3/2015/73). We thank
Anna Dyrda\l~ and Martin Gmitra for fruitful discussion. The {\it ab initio} calculations were performed on the Prometheus Computer in 
the Cyfronet Center, which is a part of the PL-Grid Infrastructure.
\section*{References}

\end{document}